\documentclass[11 pt]{article}  
 \pdfoutput=1 
\usepackage{jheppub}
\usepackage{titlesec}
\usepackage{tikz, amsmath} 
\usetikzlibrary{calc}
\usepackage{bbm}  
\usepackage{xcolor} 
\usepackage{dsfont} 
\usepackage{subfigure} 
\usepackage{hyperref}  
\usepackage{microtype}      
\usepackage{caption}
\usepackage{amsmath}
\usepackage{float} 
\usepackage{graphics}
\usepackage{psfrag} 
\usepackage{color}  
\usepackage{slashed}
\usepackage{subfigure}
\usepackage{graphicx}
\usepackage{array,multirow}
\usepackage{amsfonts} 
\usepackage{amssymb}
\usepackage{mathtools}
\usepackage{amsmath} 
\usepackage{amsfonts}
\usepackage{mathrsfs}
\usepackage{multirow}
\usepackage{feynmp}
\usepackage{array}
\usepackage{textcomp} 
\usepackage{phaistos} 
\usepackage[utf8]{inputenc}
\usepackage{pifont} 
\usepackage{pgf} 

\usepackage{subfigure}
\usepackage{multirow}
\usepackage{tabu}

\newcommand{\beq}{\begin{eqnarray}}
\newcommand{\eeq}{\end{eqnarray}}

\newcommand{\bmp}{\noindent\begin{minipage}{16cm}}
\newcommand{\emp}{\end{minipage}\vskip 7mm} 

\usepackage{dcolumn}
\usepackage{bm}
\usepackage{slashed}

\usepackage{epsfig}

\usepackage[margin=5pt, font=normalsize,labelfont=bf,justification=raggedright]{caption}

\usepackage{hyperref}

\newcommand{\bea}{\begin{eqnarray}}
\newcommand{\eea}{\end{eqnarray}}

\newcommand{\ba}{\begin{eqnarray}}
\newcommand{\ea}{\end{eqnarray}}
    
\newcommand{\Tr}{\mbox{Tr}\;}

\title{\Large Dark top partner}
\author[1]{Haiying Cai,}
\emailAdd{hcai@korea.ac.kr}
\affiliation[1]{Department of Physics, Korea University, Seoul 136-713, Korea}

\author[2,3]{Giacomo Cacciapaglia}
\emailAdd{g.cacciapaglia@ipnl.in2p3.fr}
\affiliation[2]{Institut de Physique des Deux Infinis de Lyon (IP2I), UMR5822, CNRS/IN2P3, F-69622 Villeurbanne Cedex, France}
\affiliation[3]{University of Lyon, Universit\'e Claude Bernard Lyon 1,
F-69001 Lyon, France} 

\abstract{Composite Higgs models with extended symmetries can feature mesonic dark matter candidates. In fundamental CHMs, the origin of dark parity  can be explained in the UV theory.  Combined with top partial compositeness, this leads to non-chiral  Yukawa interaction connecting  mesonic DM  with one dark  top partner and one SM  top. We examine the DM phenomenology in  SU(6)/SO(6) and SU(6)/Sp(6)  CHMs with the presence of dark top partners.   Phenomenological constraints require the mass of   top partner in even parity  to be of  the multi-TeV order.}

\begin{document}
\maketitle

\section{Introduction}
 
With the discovery of the Higgs boson at a mass of 125 GeV at the Large Hadron Collider, the particle content predicted by the Standard Model (SM) is complete. Yet, this does not close the door to the presence of New Physics. Arguably, the presence of a Dark Matter (DM) component in the present-day universe is the most compelling evidence for new physics \cite{Bertone:2018krk}, as no part of the SM can provide a particle candidate for it.

Models of new physics based on a strong confining dynamics can explain both the naturalness of the Higgs mass and the presence of Dark Matter. In fact, both the Higgs and a set of stable new particles may emerge as composite states. The naturalness of the scale is related to its dynamical origin, in the same spirit as the QCD scale. Composite  models applied to the electroweak (EW) scale are as old as the SM itself \cite{Weinberg:1975gm}, where the Higgs boson can be associated to a light pseudo-Nambu-Goldstone boson (pNGB) via vacuum misalignment \cite{Kaplan:1983fs}. Finally, an attractive mechanism to generate the top quark mass may be related to the presence of spin-1/2 operators with linear couplings to the elementary top fields in the SM. This leads to the idea of top partial compositeness \cite{Kaplan:1991dc}. This old idea has been revamped in the 2000's thanks to the discovery of holography \cite{Contino:2003ve}, based on the idea of walking dynamics \cite{Holdom:1981rm}. Realistic models have been constructed based on underlying gauge-fermion theories, leading to a limited number of combinations of gauge groups and fermion representations \cite{Barnard:2013zea,Ferretti:2013kya}. In such scenarios, a DM candidate can emerge as an additional pNGB that accompanies the Higgs boson \cite{Frigerio:2012uc}. In this direction, many symmetry breaking patterns have been examined in the literature \cite{Ballesteros:2017xeg,Balkin:2017aep,Balkin:2018tma,Cai:2018tet,Cacciapaglia:2019ixa,Ramos:2019qqa, Chala:2021ukp,Cacciapaglia:2021aex}. 

Fundamental composite dynamics models \cite{Cacciapaglia:2014uja,Cacciapaglia:2020kgq}, based on gauge-fermion underlying theories, however, limits the symmetry breaking patterns that can be realised. Depending on the nature of the fermions representations under the confining groups, we have: SU($N$)$^2$/SU($N$) for complex representations, with minimal $N=4$; SU($2N$)/Sp($2N$) for pseudo-real, with minimal $N=2$; SU($N$)/SO($N$) for real, with minimal $N=5$. In the SU(4)/Sp(4) and SU(5)/SO(5) CHMs,  although the Higgs is accompanied by additional pNGBs,  the CP-odd one decays via the topological anomaly \cite{Wess:1971yu,Witten:1983tw}.  For  the last two types of cosets, DM candidates emerge in the  extensions of  SU(6)/Sp(6) \cite{Cai:2018tet} and  SU(6)/SO(6) \cite{Cacciapaglia:2019ixa,Cai:2020njb}.  As discussed in this paper,   in the real and pseudo-real realizations  the dark parity of composite states naturally  originate  from a $\mathbb{Z}_2$ parity carried by additional fundamental fermions.

In this work, we will reexamine these two cases by including the effect of top partners in the properties of the DM candidates. Hence, the results presented here complete and complement the literature. For the first time, we give the explicit embeddings  of top partners in two SU(6) CHMs. In particular, we discuss the effect of non-chiral Yukawa interaction from the dark top partners on the DM relic density and  the direct detection constraints. The models we consider are predictive as the possible choices of partial compositeness couplings is strongly limited by the requirement of preserving the dark parity, which keep the DM pNGBs stable. The dark top partners can also be produced at hadron colliders, like the LHC, as they will typically decay into a top quark plus a DM candidate, leading to missing transverse energy signatures. It has been shown that searches for supersymmetric tops effectively cover this signature \cite{Kraml:2016eti}, with the limits on the dark top partners mainly stemming from the larger production rates as compared to the stops.

The article is organized as follows: after discussing the basic features of the models and the origin of the dark parity in Section~\ref{sec:parity}, we present the dark top partners in Section~\ref{sec:tops}. In Section~\ref{sec:EWPO} we discuss the limits on the models stemming from electroweak precision tests, before discussing the impact on DM in Section~\ref{sec:DM}. Finally, we present our conclusions in Section~\ref{sec:6}.

\section{Dark parity in Fundamental Composite Higgs Models} \label{sec:parity} 

In models with a fundamental composite dynamics, the global symmetry is broken due to the phase transition of a strong gauge dynamics. The condensation of Hyper-Color  fermions  leads to  composite  pNGBs and top partners. For the HC  fermions $\psi$ in the pseudo-real and real representations,  no dark matter candidates are allowed in the minimal cosets of  SU(4)/Sp(4) and SU(5)/SO(5) respectively.  For these two types of  realizations, a dark matter candidate can be generated by extending the content of HC fermions. We will consider here models containing the minimal set of fermions to generate a  pNGB Higgs, and enlarge with necessary number of dark fermions. As the dark fermions are not compositions of the pNGB Higgs, they can  be  odd  under a $\mathbb{Z}_2$ parity. With an appropriate embedding, the $\mathbb{Z}_2$ symmetry will be preserved after the HC fermion condensation,  as long as  the spurions of  the electroweak gauge  and Yukawa interactions are  invariant under this  parity.  Therefore, this dark parity emerging as an EFT symmetry is in fact  associated to the HC fermions in the UV theory.

In particular, the dark parity is conserved in the Wess-Zumino-Witten (WZW) topological term~\cite{Wess:1971yu, Witten:1983tw} in the EFT.  We start with the pNGBs embedded in the coset of $\mathcal{G}/\mathcal{H}$:
\begin{eqnarray} \label{eq:pNGBmatrix}
U_{\Pi}=e^{\left(i \frac{ \sqrt{2}   }{f} ~ \Pi \right)}, ~~  \text{with} ~~  \Pi = \sum_{i=1}^{dim(\mathcal{G}/\mathcal{H})}  \pi_{i} U_\alpha X^{i} U_\alpha^{-1}\,, 
\end{eqnarray}
where $U_\alpha = \exp (i \sqrt{2} \alpha  X_h)$ with $X_h$ being Higgs generator is the rotation matrix  that misaligns the  vacuum and  the generators are normalized as $\Tr X^{i} X^{j} =  \delta^{ij}$ (real) and $\Tr X^{i} X^{j} =  \frac{1}{2} \delta^{ij}$ (pseudo-real).  However unlike QCD, there is   a misalignment effect  in the composite sector.  Here we will focus on  the real and pseudo-real types of realizations, where  the vaccum along the EW symmetry breaking  direction is defined as  $\Sigma_\alpha = U_\alpha \Sigma_{EW} U_\alpha^{T}$. And in these two cases,  using the differential form approach~\cite{Kaymakcalan:1983qq}, the WZW Lagrangian can be derived to be:
\beq
\mathcal{L}_{WZW} &\supset& \frac{g_{a} g_{b} \, d_\psi }{48 \sqrt{2} \pi^2 f}  \epsilon^{\mu \nu \alpha \beta} V_{\mu \nu}^a  V^b_{\alpha \beta} \, \Big( \Tr \left[ 2 \left(T^a \, T^b    +   T^b \, T^a \right)  \Pi \right]  \nonumber \\
& &  \quad  -  \Tr \left[ T^a \, \Pi \, \Sigma_\alpha  \, (T^{b})^T  \Sigma^\dagger_\alpha + (T^{a})^T  \Sigma^\dagger_\alpha \, \Pi \, T^b  \, \Sigma_\alpha \right]  \Big)
\eeq
with $V_{\mu \nu}^a = \partial_\mu V^a_\nu - \partial _\nu V^a_\mu$,  $V_\mu^a =   W_\mu^i (i = 1,2,3) ,   B_\mu $ and $g_a$ being the corresponding gauge couplings. The  $d_\psi$ is the representation dimension of HC fermion $\psi$ in the UV gauge theory. And the exact coefficients for anomaly terms in $SU(6)/SO(6)$ CHM were first calculated in~\cite{Cacciapaglia:2019ixa}. Note that  the dark  parity needs to be a good symmetry after the vacuum misalignment, i.e. $\Omega_{DM} U_\alpha \Omega_{DM} = U _\alpha$.  Hence under the $\mathbb{Z}_2$ parity operation,  the rotated $\Pi$ matrix transforms as $\Omega_{DM} \Pi (\pi_{odd}) \Omega_{DM} = \Pi (- \pi_{odd})$, where only the odd parity  pNGBs change to be in the opposite sign.  In addition, because the $W_\mu^i, B_\mu$ gauge bosons are even fields under this parity,   the dark parity will prevent the  odd parity pNGB from decaying via WZW terms.

For the complex realization of CHM,   in principal one can also define the dark parity  in terms of  the  HC fermion operators. While we can  understand  the  $\mathbb{Z}_2$ parity for the  breaking pattern $SU(N)_l\times SU(N)_r /SU(N)_v$  from an EFT perspective. In  this  scenario,   the dark parity  is equivalent to exchange the flavor groups $SU(N)_l \leftrightarrow SU(N)_r$. Defining $X^i$ as the SU(N) generators,  the parity with the property of  $\Omega_{DM} X^{i, T} \Omega_{DM}^\dag = \pm X^i$ is in fact the EW preserving vacuum that breaks SU(N) $\to$ SO(N) or Sp(N).  As a concrete example, the odd parity pNGBs in $SU(5)^2/SU(5)$ CHM decompose as $(2,2)\oplus (1,3)\oplus (3,1)$ under the EW $SU(2)_L \times SU(2)_R$ symmetry.  Due to this parity,  the WZW term structure in $SU(N)^2/SU(N)$ is the same as in  SU(N)/SO(N) or SU(N)/Sp(N). 
 In the following, we will mainly  illustrate the  features of the resulting two minimal models with dark matter: SU(6)/SO(6)  and SU(6)/Sp(6). 

\subsection{SU(6)/SO(6)}

The minimal set of fundamental fermions consists of two doublets with opposite hypercharge and a singlet. Hence, we include a second singlet, odd under the dark parity. The fermion content is illustrated in Table~\ref{tab:real}.

\begin{table}[htb]
\centering
\begin{tabular}{|l||c|c||c|c|}
\hline
Real & SU(2)$_L $ & U(1)$_Y$ & SU(2)$_R$ & $\mathbb{Z}_2$ \\ \hline
$\psi_{D1}$ & $\bf 2$ & $\bf 1/2$ & \multirow{2}{*}{$2$} & $+$ \\
$\psi_{D2}$ & $\bf 2$ & $\bf -1/2$ & & $+$ \\
$\psi_{S1}$ & $\bf 1$ & $\bf 0$ & $1$ & $+$ \\
$\psi_{S2}$ & $\bf 1$ & $\bf 0$ & $1$ & $-$ \\
\hline
\end{tabular}
\caption{\label{tab:real} Hyper-Color fermions in the case of real realization, and their quantum numbers. The two fermions $\psi_{D1}$ and $\psi_{D2}$ transform as bi-doublet under the $SU(2)_L \times SU(2)_R$ with $U(1)_Y \subset  SU(2)_R$.The last two columns indicate the global symmetries, i.e. the custodial SU(2)$_R$ and the dark parity.}
\end{table}

Upon condensation, the symmetry breaking pattern SU(6)/SO(6) emerges, leading to 20 pNGBs. A subset of them will be odd under the dark $\mathbb{Z}_2$, hence playing the role of Dark Matter candidates~\cite{Cacciapaglia:2019ixa, Cai:2020njb}.
We summarise here their main features, in preparation for the inclusion of top partners. The $SU(2)_L\times SU(2)_R$ generators that are invariant  under $\mathbb{Z}_2$ are:
\begin{eqnarray}
S_{L}^{1,2,3} =    \frac{1}{2} 
\left(
\begin{array}{c|c}
\mathbbm{1}_2  \otimes  \sigma_i   & \\ \hline
& 0_{2} \end{array} \right) \,, 
\qquad 
S_{R}^{1,2,3}=   \frac{1}{2} 
\left(
\begin{array}{c|c}
\sigma_i \otimes \mathbbm{1}_2  & \\ \hline
& 0_{2} \end{array} \right)
\end{eqnarray}
that  satisfy the condition $S_{L/R}^i \Sigma_{EW} + \Sigma_{EW} S_{L/R}^{i T} =0$, with the   EW  preserving vacuum  to be:
\begin{eqnarray}
\Sigma_{\rm EW}  =\left(
\begin{array}{cc|cc}
& i \sigma_2 &  \\
-i \sigma_2 &  & \\ \hline
& & \mathbbm{1}_{2} 
\end{array} \right)\,.
\end{eqnarray}
The $\Pi$ matrix in the SU(6)/SO(6) CHM can be written as 
\begin{eqnarray}
2 ~\Pi = \left(\begin{array}{cccc}
\varphi + \frac{\eta _1}{\sqrt{3}} \mathbbm{1}_{2} &  \Lambda    &  \sqrt{2} {H}_{1}  &  \sqrt{2} {H}_{2} \\
\Lambda^{\dagger}     & - \varphi + \frac{\eta _1}{\sqrt{3}} \mathbbm{1}_{2} & -\sqrt{2}  \widetilde{{H}_{1}} & - \sqrt{2} \widetilde{{H}_{2}} \\         
\sqrt{2} {H}_{1} ^{\dagger} & - \sqrt{2}  \widetilde{{H}_{1}}^{\dagger} &2 \left( \frac{\eta _3}{\sqrt{2}}-\frac{\eta
	_1}{\sqrt{3}}  \right) & \sqrt{2} \eta _2 \\
\sqrt{2} {H}_{2} ^{\dagger} & - \sqrt{2}  \widetilde{{H}_{2}}^{\dagger}  &  \sqrt{2} \eta _2 & -2 \left( \frac{\eta _3}{\sqrt{2}}+\frac{\eta
	_1}{\sqrt{3}}  \right)\\
\end{array} \right)\,,
\end{eqnarray}
where the  columns and rows correspond to the fermions in Table~\ref{tab:real}. We can see that  the first  doublet $H_1$ is composed of   HC fermions $(\psi_{D1}, \psi_{D2})$ and $\psi_{S1}$ in even parity. The matrices $\varphi$ and $\Lambda$ represent the triplets, transforming as a bi-triplet of the custodial SU(2)$_L \times$SU(2)$_R$ symmetry, like in the SU(5)/SO(5) model~\cite{Agugliaro:2018vsu}. The two Higgs doublets read:
\begin{eqnarray}
H_1 = \frac{1}{\sqrt{2}} \left( \begin{array}{c} G_2 + i G_1 \\ h - i G_3  \end{array} \right) \,, \qquad       H_2 = \left( \begin{array}{c} H_+ \\ \frac{H_0 + i A_0 }{\sqrt{2}} 
\end{array} \right) \,, \qquad  \widetilde{H}_{1,2} = i \sigma_2 H_{1,2}^*\,, \label{eq:SO6doublet}
\end{eqnarray}
where the second doublet stems from  $(\psi_{D1}, \psi_{D2})$ and the odd parity singlet $\psi_{S2}$, while $\eta_i, i = 1,2,3 $ are gauge singlets. Note that the Goldstone bosons $G_{1,2,3}$ inside  the first doublet $H_1$  are eaten by $W^{1,2}_{\mu}$ and $Z_\mu$ respectively~\footnote{The broken generators for the eaten Goldstone bosons $G_{i}, i= 1,2,3$ are slightly adjusted compared with the ones in \cite{Cacciapaglia:2019ixa} in order to get a universal  formula  in Eq.(\ref{eq:Emu}-\ref{eq:dmu}).}.
 
Following the Hyper-Color fermions in Table~\ref{tab:real},   the dark parity in the EFT is defined as
\begin{eqnarray}
\Omega_{\rm DM} = \left( \begin{array}{ccc}
\mathbbm{1}_{2} &   &  \\
& \mathbbm{1}_{2} &   \\
&   & \sigma_{3}  \\
\end{array} \right) \,.
\end{eqnarray}
that  acts on the pNGB matrix $\Sigma = U_\Pi \Sigma_\alpha U^T_{\Pi}$  in the following way \cite{Cacciapaglia:2019ixa}:
\begin{eqnarray}
\Omega_{\rm DM} \Sigma (H_2, \eta_2)\Omega_{\rm DM} =  \Sigma (- H_2, -\eta_2)\,,
\end{eqnarray}
hence the $\mathbb{Z}_2$--odd pNGBs are the second doublet $H_2$ and the singlet $\eta_2$. 
We will discuss the dark parity of the top partners of this model in the next section. Since the top partners normally are much heavier than the pNGBs,  they mainly participate in the pNGB  DM production as  calculated in Section~\ref{sec:DM}.

\subsection{SU(6)/Sp(6)}
 
For pseudo-real realization, the minimal HC fermion content consists of one $SU(2)_L$ doublet $\psi_L$ and two singlets with opposite hypercharge, forming a doublet of the global SU(2)$_{R1}$.  When one extends this type of  model, the dark sector needs to  consist of two additional $\mathbb{Z}_2$--odd fermions.  In order to avoid pNGBs with semi-integer charges and gauge anomalies, these states must have opposite semi-integer hypercharges. Here, we follow the minimal choice as shown in Table~\ref{tab:pseudo}. The odd and even singlets can be organized as doublets of two custodial SU(2)$_R$ symmetries, where the first acts on the SM-like Higgs, while the second acts on the odd doublet.
\begin{table}[htb]
\centering
\begin{tabular}{|l||c|c||c|c|c|c|}
\hline
Pseudo-real & SU(2)$_1$ & U(1)$_Y$ & SU(2)$_{R1}$ & SU(2)$_{R2}$ &  $\mathbb{Z}_2$ \\ \hline
$\psi_{L}$ & $\bf 2$ & $\bf 0$ & $1$ & $1$ & $+$ \\
$\psi_{R1}$ & $\bf 1$ & $\bf \mp 1/2$ & $2$ & $1$ & $+$ \\
$\psi_{R2}$ & $\bf 1$ & $\bf \mp 1/2$ & $1$ & $2$ & $-$ \\
\hline
\end{tabular}
\caption{\label{tab:pseudo} Hyper-Color fermions in the case of pseudo-real realization, and their quantum numbers. The last two columns indicate the global symmetries, i.e. the two custodial SU(2)$_{R1}\times$SU(2)$_{R2}$ and the dark parity.}
\end{table}

The pNGB in the $SU(6)/Sp(6)$ CHM was investigated in~\cite{Cai:2018tet} and the  generators of  $SU(2)_1 \times SU(2)_{R1} \times SU(2)_{R2}$ are:
\begin{equation}
S^{1,2,3} = \frac{1}{2}\begin{pmatrix}
\sigma_i & 0 & 0\\
0 & 0 & 0\\
0 & 0 & 0
\end{pmatrix}\,,
\quad 
S^{4,5,6} = \frac{1}{2}\begin{pmatrix}
0 & 0 & 0\\
0 & -\sigma_i^T & 0\\
0 & 0 & 0
\end{pmatrix}\,,
\quad 
S^{7,8,9} = \frac{1}{2}\begin{pmatrix}
0 & 0 & 0\\
0 & 0 & 0\\
0 & 0 & -\sigma_i^T
\end{pmatrix} \,.
\end{equation}
They are left unbroken by the condensate
\beq
\Sigma_{EW} = \left(
\begin{array}{ccc}
 i \sigma _2  & 0 & 0 \\
 0 & -i \sigma _2  & 0 \\
 0 & 0 & -i \sigma _2  \\
\end{array}
\right)
\eeq
that obviously preserves the dark parity. In Table~\ref{tab:pseudo},  the gauged $SU(2)_L$ group is the first flavor $SU(2)_1$ subgroup  and the gauged hypercharge $U(1)_Y$ is defined as~\footnote{The other option is to  gauge  $SU(2)_1 + SU(2)_{R2}$ as   $SU(2)_L$  and let   $Y = T^3_{R1}$. In such a case,   the EW quantum numbers of  pNGBs will change, but the global symmetry breaking pattern  remains  the same.}:
\begin{equation}
    Y = T^3_{R1} + T^3_{R2}\,,
\end{equation}
i.e. the sum of the two diagonal generators of the global SU(2)$_R$ symmetries. Note, however, that the abelian gauging still leaves two U(1) global symmetries unbroken, U(1)$_{R1}\times$U(1)$_{R2}$. The first is broken by the Higgs vacuum expectation value, while the second remains unbroken. Hence for the EW gauge group embedding  in Table~\ref{tab:pseudo}, the discrete $\mathbb{Z}_2$ parity  is actually  enhanced  to a dark U(1) symmetry protecting the DM. 

The model in Table~\ref{tab:pseudo} generates the coset SU(6)/Sp(6), which has 14 pNGBs, which can be expressed as in Eq.~\eqref{eq:pNGBmatrix}, with
\beq
2 \, \Pi =  \left(
\begin{array}{cccc}
 \frac{1}{\sqrt{6}} \left(\sqrt{3} \eta _1+  \eta _2\right) \mathbbm{1}_{2} &  H_1 &  H_2 \\
H_1^\dag &- \frac{1}{\sqrt{6}} \left(\sqrt{3} \eta _1-  \eta _2\right) \mathbbm{1}_{2}  & \Phi \\
 H_2^\dag & \Phi^\dag & - \sqrt{\frac{2}{3}} \eta _2
\end{array}
\right)\,.
\eeq
with  the bi-doublets explicitly to be
\beq
H_1 =   \left(
\begin{array}{cc}
\frac{- G_1 + i G_2}{\sqrt{2}}  &  \frac{G_3-i h }{\sqrt{2}} \\
\frac{G_3 + i h}{\sqrt{2}} & \frac{G_1+i G_2}{\sqrt{2}}
 \end{array} \right)  \,, ~~  H_2 = \left(
\begin{array}{cc}
 H_+ &  \frac{A_0-i H_0}{\sqrt{2}} \\
 \frac{A_0+i H_0}{\sqrt{2}} & - H_-
 \end{array} \right) \,, ~~ \Phi = \left(\begin{array}{cc}
 \frac{ \eta _4 + i \eta_3}{ \sqrt{2}} & \eta _- \\
 - \eta _+ & \frac{\eta _4-i \eta _3}{\sqrt{2}}
\end{array} 
\right) . \label{eq:Sp6doublet}
\eeq
%
Here we can recognize the components of the SM Higgs doublet $H_1$ and of a second doublet $H_2$,  like in the previous model.  Furthermore, $\big(\eta_\pm \,, \frac{\eta_3 +i  \eta_4}{\sqrt{2}}\big)$ in $\Phi$ are singlets that form a bi-doublet of SU(2)$_{R1}\times$SU(2)$_{R2}$. Finally, $\eta_1$ and $\eta_2$ are singlets. Note that in terms of the HC fermions condensation,  $H_2$ is from  $\psi_{L} \psi_{R_2}$ and  $\Phi$  is from $\psi_{R1} \psi_{R2}$.  Hence in the EFT, the dark parity  reads:
\beq
\Omega_{DM} = \left( \begin{array}{c c}
\mathbbm{1}_4 & 0 \\
0 & - \mathbbm{1}_{2} 
\end{array} \right) \,,
\eeq
under which the odd states are the second  doublet $H_2$ and the singlets $\eta_{\pm}, \frac{1}{\sqrt{2}}(\eta_3 +i  \eta_4)$, i.e. the only pNGBs that transform as doublets under SU(2)$_{R2}$.  Due to the enhanced $U(1)$ symmetry, the $ \frac{1}{\sqrt{2}}(\eta_3 +i  \eta_4)$ will be a complex DM. And like in the previous case, $\Omega_{DM}$ will also be used to classify the top partners as discussed in Section~{\ref{sec:tops}}.

\section{Dark top partners and partial compositeness} \label{sec:tops}

Top partners are spin-1/2 composite state that generate the top mass via linear mixing with the elementary top fields~\cite{Kaplan:1991dc, Contino:2006nn}. Hence, the mass eigenstate corresponding to the physical top is a mixture of elementary and composite states. This mixing plays a crucial role in determining both the vacuum misalignment and the properties of the Higgs boson and  preserve  dark parity in the extended models.  Here we will  study the  compositions of top partners as condensation of  HC fermions and provide their embeddings in the effective Lagrangian.  In particular, the  properties of the odd top partners are investigated, which were not considered before.

In fundamental composite models, the top partners are determined by the microscopic dynamics of the confining interactions. In general terms, top partners are resonances associated to some spin-1/2 operators made of the HC fermions. As such, they transform coherently under the unbroken global symmetry in the confined sector. In the models of Refs~\cite{Barnard:2013zea, Ferretti:2013kya, Ferretti:2016upr}, such operators are built of two species of fermions: the electroweak ones $\psi$, which we introduced in the previous section, and a new species $\chi$, which carries QCD charges and the appropriate hypercharges. The $\chi$ sector will be extended in order to accommodate the top partners and  we will follow the nomenclature of models introduced in Ref.~\cite{Belyaev:2016ftv}. 

For the two SU(6) CHMs,  the top partners  correspond to the operators of $\psi \psi \chi$, $\bar \psi \bar \psi \chi$ and $\psi \bar{\psi} \bar{\chi}$ that are  singlets under the HC gauge group.  In Table~\ref{tab:tops}, the top partners are classified according to their transformation property under the unbroken flavor groups of $SO(6)\times SO(6)$ or  $Sp(6)\times SO(6)$, with the second flavor group related to the QCD  sector.  Depending on  $\mathcal{G_{HC}}$,  the first two  $\psi \psi \chi$, $\bar \psi \bar \psi \chi$  will either be in the  2-index symmetric $ S_2$  or antisymmetric $ A_2$ representations of  the flavor  group SU(6).  While  the third composition  $\psi \bar{\psi} \bar{\chi}$ contains the adjoint and singlet representations of the global SU(6).  Under the unbroken group, the adjoint decomposes into $S_2$ and $A_2$  of the SO(6) or Sp(6) global symmetries in the EW sector. 
In the following we will study top partners belonging to the anti-symmetric, $\Psi_A$,  leaving the others for future studies.
We only  accompany a singlet $\Psi_1$ with $\Psi_A$  in the $SU(6)/Sp(6)$ model, as  $\not d \Psi_1$ is  symmetric in SU(N)/SO(N). Their effective Lagrangian can be written as \cite{Cai:2022zqu}:  
\beq
\mathcal{L}_{\rm composite} &=& tr \left[ \bar{\Psi}_A\ i \not{D}   \Psi_A \right] - M_A \ tr \left[ \bar{\Psi}_A  \Psi_A \right] + tr \left[ \bar{\Psi}_1\ i \not{D}\ \Psi_1\right] - M_1 \ tr \left[\bar{\Psi}_1  \Psi_1\right]  \nonumber \\
&+&   \kappa' \,  tr \left[ \bar{\Psi}_A \not{d}\  \Psi_A \right]  + \kappa \, \left( tr \left[ \bar{\Psi}_A \not{d}\  \Psi_1 \right]  + \mbox{h.c.} \right)  \,. \label{eq:Lcomposite}
\eeq
where the two masses $M_A$ and $M_1$  are naturally expected to be a few times the pNGB decay constant $f$.  The covariant derivative reads
\beq
D_\mu  = ( \partial_\mu  - i E_\mu - i g_1 X  B_\mu  - ig_s G^a_\mu \lambda_a  )\,.
\eeq
with $X$ to be the hypercharge carried by the $\chi$ and  including the misalignment effect,
\beq
E_\mu &=& \sum_i^3 \left( g_2 W_\mu^i  T_L^i + g_1 B_\mu (T_{R1}^3+ T_{R2}^3) \right) - s^2_{\frac{\alpha}{2}}  \sum_i^3 \left( g_2 W_\mu^i -g_1 B_\mu \delta^{i 3 }\right) \left(T_L^i -T_{R1}^i \right) + \cdots  \,, \label{eq:Emu}
\eeq
The dots indicate higher order terms with the presence of pNGBs. The $\kappa$ terms contain the Maurer-Cartan  form aligned with the broken generators, which reads
 \beq \label{eq:dmu}
d_\mu &=& -  \frac{\sqrt{2 }}{f} \partial_\mu  \Pi +    \frac{s_\alpha}{\sqrt{2}} \sum_{i=1}^3 (g_2 W^i_\mu - g_1 B_\mu \delta^{i 3})  X_G^i + \cdots. \,,  \label{eq:dmu}
\eeq
with $X_{G}^i$ being the generators for  the eaten Goldstone bosons shown in Eq.(\ref{eq:SO6doublet}) and Eq.(\ref{eq:Sp6doublet}). As discussed in~\cite{Cai:2022zqu},  the CCWZ forms $E_\mu$ and $d_\mu$ are universal at the leading order for a generic CHM.  The covariant derivative and the $\kappa$ terms parameterize the $1/f $ suppressed interactions with the pNGBs, as well as corrections to the EW gauge couplings due to misalignment. The effect of the latter on EW precision tests has been first discussed in Ref.~\cite{Cai:2022zqu} in a minimal model. Note that the self-conjugate $\kappa'$ term only appears in the extended models thanks to the unbroken SO(6) or Sp(6) symmetries, but it is absent in the minimal models.

The linear mixing with the elementary top fields crucially depends on the properties of the composite operator generating the top partners. While the elementary fields need to be embedded in an incomplete representation of SU(6).  In this paper we choose the spurions to be in  the adjoint.  The matching is performed by dressing the pNGB matrix $U_\Pi$ in Eq.~(\ref{eq:pNGBmatrix}), leading to the  terms of partial compositeness (PC):
\beq \label{eq:Lmix}
\mathcal{L}_{\rm mix} &=&  y_{L} f \ tr \left[ D_{L}^\dag \gamma_0 U_{\Pi}  \Psi_A  \Sigma_\alpha^* U_{\Pi}^\dag \right]  +  y_{R} f \  tr \left[ D_{R}^\dag \gamma_0 U_{\Pi}  \Psi_A  \Sigma_\alpha^* U_{\Pi}^\dag \right] + \text{h.c.}  \, \label{eq:PC}
\eeq
where $\Sigma_\alpha$ is the misaligned vacuum, and $D_L$ and $D_R$ are the spurions of the left-handed doublet and the right-handed top, respectively. Their explicit form depends on the model and will be discussed below. Finally, $y_L$ and $y_R$ parameterize the strength of the linear mixing of the top partners with the elementary fields. The Lagrangian in Eq.(\ref{eq:PC}) gives rise to the mass matrix of  spin-1/2 states  and determine the Yukawa couplings at the higher order,  that connect the top partners to one SM field and one pNGB.  The couplings involving the dark top partners relevant to DM production are listed in Appendix~\ref{Appendix2}.

\begin{table}[tb]
\begin{center}
  {\tabulinesep= 1.5 mm
\begin{tabu}{|c|c|c|c|c|}
\hline
   &\multicolumn{2}{|c|}{SO(6)$\times$SO(6)} & \multicolumn{2}{|c|}{Sp(6)$\times$SO(6)}  \\
   \hline \hline
 $\mathcal{G}_{HC}$ &  SO(7)  &  SO(9)    & SO(11)  & Sp(4) \\
\hline  \hline
$ \psi \psi \chi $ or  $\bar{\psi} \bar{\psi} \chi$  &  $(\bf{15}, \bf{6})$ & $(\bf{1}, \bf{6})$ & $(\bf{21}, \bf{6})$ & $(\bf{1}, \bf{6}) $   \\
                       &           &      $(\bf{20}, \bf{6})$   &    & $(\bf{14}, \bf{6})$   \\
\hline
$\psi \bar{\psi} \bar{\chi} $  &   \multicolumn{2}{|c|}{$(\bf{1}, \bf{6})$}  &  \multicolumn{2}{|c|}{$(\bf{1}, \bf{6})$}  \\
\hline
   $\psi \bar{\psi} \bar{\chi}$    & \multicolumn{2}{|c|}{$(\bf{15}, \bf{6}) \oplus (\bf{20}, \bf{6})$ }  & \multicolumn{2}{|c|}{$(\bf{14}, \bf{6})\oplus (\bf{21}, \bf{6}) $}  \\
\hline 
\end{tabu}}
\caption{Top partners  in two SU(6) CHMs are classified  with respect to  the unbroken flavor subgroups and  $\mathcal{G}_{HC}$ is the Hyper-Color  gauge group. }
\label{tab:tops}
\end{center}
\end{table}

\subsection{SU(6)/SO(6)}
  
For the SU(6)/SO(6) coset, the desired top partners are obtained for models with confining gauge groups SO(7) and SO(9), where the $\psi$ fermions are in the spinorial  and the $\chi$ fermions in the fundamental representation~\cite{Belyaev:2016ftv}.  In  Table~\ref{tab:tops},  the combination $\psi \psi$ (or $\bar \psi \bar \psi $)  inside a top partner   $\psi \psi \chi$ (or $\bar \psi \bar \psi \chi$) is  a fundamental representation in SO(7) or SO(9) with specific symmetry as a decomposition of  tensor square.  
 
For $\mathcal{G}_{HC} = SO(7)$,  the $\psi \psi \chi$  gives rise to a  two-index anti-symmetric representation  $\bf 15$ in SO(6).  Under the custodial SU(2)$_L \times$SU(2)$_R$, it decomposes as
 \begin{eqnarray}
{\mathbf 15} &\to&   (2,2)\oplus (2,2)\oplus (1,3)\oplus (3,1) \oplus (1,1)\,.
\end{eqnarray}
The top partners carrying a single $\mathbb{Z}_2$-odd HC fermion will be odd under the operation of dark parity. Hence, the top partner field can be written as
\beq
 \Psi_{15} &=& \Psi_{(3,1)} +  \Psi_{(1,3)} +  \Psi_{(2,2)} + \tilde{\Psi}_{(2,2)} + \tilde{\Psi}_{(1,1)}\,, 
\eeq
where the various components are expressed in the following form:
\beq
\Psi_{(3,1)} = i \left(
\begin{array}{cccccc}
 0 & 0 & -\frac{X_{\frac{5}{3}}}{\sqrt{2}} & \frac{X_{\frac{2}{3}}}{2} & 0 & 0 \\ 
 0 & 0 & \frac{X_{\frac{2}{3}}}{2} & \frac{X_{-\frac{1}{3}}}{\sqrt{2}} & 0 & 0 \\
 \frac{X_{\frac{5}{3}}}{\sqrt{2}} & -\frac{X_{\frac{2}{3}}}{2} & 0 & 0 & 0 & 0 \\
 -\frac{X_{\frac{2}{3}}}{2} & -\frac{X_{-\frac{1}{3}}}{\sqrt{2}} & 0 & 0 & 0 & 0 \\
 0 & 0 & 0 & 0 & 0 & 0 \\
 0 & 0 & 0 & 0 & 0 & 0 \\
\end{array}
\right) \,,
 \quad 
 \Psi_{(1,3)} = i  \left(
\begin{array}{cccccc}
 0 & -\frac{Y_{\frac{5}{3}}}{\sqrt{2}} & 0 & \frac{Y_{\frac{2}{3}}}{2} & 0 & 0 \\
 \frac{Y_{\frac{5}{3}}}{\sqrt{2}} & 0 & -\frac{Y_{\frac{2}{3}}}{2} & 0 & 0 & 0 \\
 0 & \frac{Y_{\frac{2}{3}}}{2} & 0 & \frac{Y_{-\frac{1}{3}}}{\sqrt{2}} & 0 & 0 \\
 -\frac{Y_{\frac{2}{3}}}{2} & 0 & -\frac{Y_{-\frac{1}{3}}}{\sqrt{2}} & 0 & 0 & 0 \\
 0 & 0 & 0 & 0 & 0 & 0 \\
 0 & 0 & 0 & 0 & 0 & 0 \\
\end{array}
\right)\,, \nonumber
 \eeq
 
 \beq
 \Psi_{(2,2)} =   \frac{1 }{\sqrt{2}} \left(
\begin{array}{cccccc}
 0 & 0 & 0 & 0 & X & 0 \\
 0 & 0 & 0 & 0 & T_X & 0 \\
 0 & 0 & 0 & 0 & T & 0 \\
 0 & 0 & 0 & 0 & B & 0 \\
 - X & -T_X & - T & - B & 0 & 0
   \\
 0 & 0 & 0 & 0 & 0 & 0 \\
\end{array}
\right) \,, \quad 
\tilde{\Psi}_{(2,2)} = \frac{1 }{\sqrt{2}} \left(
\begin{array}{cccccc}
 0 & 0 & 0 & 0 & 0 & \tilde{X} \\
 0 & 0 & 0 & 0 & 0 &  \tilde{T}_X \\
 0 & 0 & 0 & 0 & 0 & \tilde{T}\\
 0 & 0 & 0 & 0 & 0 & \tilde{B} \\
 0 & 0 & 0 & 0 & 0 & 0 \\
 - \tilde{X} & -\tilde{T}_X& -\tilde{T} &
   - \tilde{B} & 0 & 0 \\
\end{array}
\right)\,, \nonumber
 \eeq
 
\beq
\tilde{\Psi}_{(1,1)} =   \frac{\tilde{T}_1 }{\sqrt{2}}  \left(
\begin{array}{ccc}
0_2 & &  \\
& 0_2 &  \\
&  & - \sigma_2   \\
\end{array} \right)\,,
 \eeq
where  the tilde fields satisfy $\Omega_{DM} \, \tilde \Psi  \, \Omega_{DM} = - \tilde \Psi $, i.e. the bi-doublet $(\tilde{T}, \tilde{B}, \tilde{T}_X, \tilde{X})$ and the singlet $\tilde{T}_1$ are $Z_2$-odd.  And  the other even states can  mix with the top fields, according to the chosen SM spurions.
 
In order to conserve the DM parity,  the elementary top spurions  needs to be even under the DM parity $\Omega_{DM} \, D_{L,R}\,  \Omega_{DM} = D_{L,R} $.  Under this condition,  the adjoint representation of SU(6) allows 2 possibilities for the left-handed doublet and 3 for the right-handed singlet:  
 \begin{equation} \label{eq:DL}
\begin{array}{ccc}
D_{L,A}^{1} =  \left(
\begin{array}{cccc|cc}
 &  &  &  & 0 & 0 \\
 &  &  &  & 0 & 0 \\
 &  &  &  & \frac{t_{L}}{\sqrt{2}} & 0 \\
&  &  &  & \frac{b_{L}}{\sqrt{2}} & 0 \\ \hline
-\frac{b_{L}}{\sqrt{2}} & \frac{t_{L}}{\sqrt{2}} & 0 & 0 &  &  \\
0 & 0 & 0 & 0 &  & 
\end{array} \right)\,,
&\phantom{xx}&
D_{L,S}^{2} =  \left(
\begin{array}{cccc|cc}
&  &  &  & 0 & 0 \\
&  & &  & 0 & 0 \\
&  &  &  & \frac{t_{L}}{\sqrt{2}} & 0 \\
&  &  &  &  \frac{b_{L}}{\sqrt{2}} & 0 \\ \hline
 \frac{b_{L}}{\sqrt{2}} & -  \frac{t_{L}}{\sqrt{2}} & 0 & 0 &  &  \\
0 & 0 & 0 & 0 & & 
\end{array} \right)\,, 
\end{array}
\end{equation}
and
\begin{equation}\label{eq:DR}
\begin{array}{c}
D_{R,S}^{1} = \frac{i}{2 \sqrt{3}}  t_{R} \left(
\begin{array}{ccc}
 \mathbbm{1}_2 &  &  \\
 & \mathbbm{1}_2 &  \\
  &  & - 2 \mathbbm{1}_2 
\end{array} \right)\,, \\ \phantom{xx} \\
D_{R,S}^{2} =  \frac{i}{\sqrt{2}}   t_{R} \left(
\begin{array}{ccc}
0_2 & & \\
 & 0_2 &  \\ &  & \sigma_3
\end{array} \right)\,, 
\qquad 
D_{R,A}^{3} =   \frac{i}{2}  t_{R} \left(
\begin{array}{ccc}
 \mathbbm{1}_2 &  &    \\
&  - \mathbbm{1}_2 &    \\
  &  &  0_{2}
\end{array} \right)\,. \\
\end{array}
\end{equation}
In the above, the ``A'' and ``S'' subscript indicate if the spurion is in the anti-symmetric or symmetric part.
Therefore the top and bottom spurions appearing in Eq.~\eqref{eq:Lmix} are:
 \begin{eqnarray}
D_{L} &=& Q_{A1} D_{L,A}^{1} + Q_{S2} D_{L,S}^{2} \,,\\
D_{R} &=& R_{S1} D_{R,S}^{1} + R_{S2} D_{R,S}^{2}  + R_{A3} D_{A,S}^{3}  \,.
\end{eqnarray}
Another constraint stems from avoiding tadpoles for the pNGBs other than $h_0$, as their presence would lead to a spontaneous violation of custodial symmetry. The following conditions need to be satisfied~\cite{Cacciapaglia:2019ixa}:
\beq
&& Q_{A1} Q_{S2}^* - Q_{S2} Q_{A1}^* =0 \,, \quad Q_{A1} Q_{S2}^* + Q_{S2} Q_{A1}^* =0 \nonumber \\ 
&&  R_{A3} (\sqrt{3} R_{S1} -\sqrt{2} R_{S2})^* + R_{A3}^* (\sqrt{3} R_{S1} -\sqrt{2} R_{S2})  =0\,.
\eeq
that will  be satisfied  given that  $t_L$ and $t_R $ are  paired in opposite parts (A-S or S-A) of  spurions. In fact,  the adjoint is  a realistic choice to allow for a tadpole free potential. 
Expanding \eqref{eq:Lmix} to the leading order, we obtain in the gauge basis:
\beq \label{eq:LmixSO}
\mathcal{L}_{mix} &=&  \frac{ y_L f}{\sqrt{2}} \bar t_L  \left(Q_{A1} \left(\cos (2 \alpha ) \left(T-T_X\right)+\cos (\alpha )
   \left(T_X+T\right)\right)- \sin (\alpha ) Q_{S2}
   \left(X_{\frac{2}{3}}-Y_{\frac{2}{3}}\right)\right)  \nonumber \\    
&+& \frac{y_R f}{4}   \bar t_R \left(4 R_{A3} \left(X_{\frac{2}{3}}\sin^2\frac{\alpha}{2} +Y_{\frac{2}{3}} \cos^2\frac{\alpha}{2} \right)+ \sin (2 \alpha ) \left(T-T_X\right)
   \left(\sqrt{3} R_{S1}-\sqrt{2} R_{S2}\right)\right) \nonumber \\
   &+ & y_L f \bar{b}_L \left(\sqrt{2} B \cos (\alpha ) Q_{A1}- \sin (\alpha ) Q_{S2}
   \left(X_{-\frac{1}{3}}-Y_{-\frac{1}{3}}\right)\right) + \cdots
\eeq
Note that the spurion choice fixes the structure of DM Yukawa interaction (ref Appendix~\ref{Appendix2}) and  the mixing pattern for the even states that  has impact on the  EW precision observables.  Diagonalizing the  resulting mass matrix,  the top quark mass is given by:
\beq
m_t =  -\frac{f^2 M \sin \alpha  \left(\sqrt{2} Q_{S2} R_{A3}+Q_{A1} \left(\sqrt{6} R_{S1}- 2
   R_{S2}\right)\right) y_L y_R}{2 \sqrt{M^2+2 f^2 Q_{A1}^2 y_L^2} \sqrt{M^2+f^2 R_{A3}^2 y_R^2}} \,. 
\eeq

\subsection{SU(6)/Sp(6)}

A similar analysis can be conducted for the pseudo-real case. This scenario is realized by the confining gauge symmetry Sp(4) with $\psi$ in the fundamental and $\chi$ in the two-index anti-symmetric and with SO(11) with $\psi$ in the spinorial and $\chi$ in the fundamental \cite{Belyaev:2016ftv}.

For $\mathcal{G}_{HC} = Sp(4)$, the traceless part of $\psi \psi \chi$ corresponds to a two-index anti-symmetric  $\bf 14$ in $Sp(6)$, which decompose  under SU(2)$_L\times$SU(2)$_{R1}\times$SU(2)$_{R2}$ as:
\beq
{\bf 14} \rightarrow (2,2,1) \oplus (2,1,2) \oplus (1,2,2) \oplus (1,1,1) \oplus (1,1,1)\,.
\eeq
Explicitly, the top partners can be written in the following form
\beq
\Psi_{14} = \left ( \begin{array}{ccc} 
\frac{i \sigma_2}{2} (T_1 + \frac{1}{\sqrt{3}} T_2)  & \psi_{(2,2, 1)}  & \tilde{\psi}_{(2, 1, 2)} \\
-\psi_{(2,2,1)}^T &  \frac{i \sigma_2}{2} (T_1 -  \frac{1}{\sqrt{3}} T_2) \  & \tilde{\psi}_{(1,2,2)} \\
- \tilde{\psi}^T_{(2,1,2)} & - \tilde{\psi}^T_{(1,2, 2)} &  \frac{i \sigma_2}{\sqrt{3}} T_2 
\end{array} 
\right) \,, 
\eeq
with $T_1$ and $T_2$ being singlets,  and three bi-doublets under the global symmetry $SU(2)^3$,
\beq
\psi_{(2,2, 1)} = \frac{1}{\sqrt{2}} \left( \begin{array}{cc}
T & X \\
B & T_X
\end{array} \right) \,, ~
\tilde{\psi}_{(2,1, 2)} =   \frac{1}{\sqrt{2}} \left( \begin{array}{cc}
\tilde{T} & \tilde{X} \\
\tilde{B} & \tilde{T}_X
\end{array} \right) \,, ~
\tilde{\psi}_{(1,2, 2)} =  \frac{1}{\sqrt{2}} \left( \begin{array}{cc}
\tilde{X}_{-1/3} & \tilde{X}_{2/3} \\
\tilde{Y}_{2/3} & \tilde{X}_{5/3}
\end{array} \right)
\eeq
where $\tilde \psi_{(2,1, 2)}$ and $\tilde \psi_{(1,2, 2)}$,  that contain one $\psi_{R2}$ in the condensations,  are  $\Omega_{DM}$-odd.
And the singlet top partner involving  in the interaction of  $Tr[\Psi_{14} \not d \Psi_1]$ is:
\beq
\Psi_1 =   \frac{T_3}{\sqrt{6}}   \left(
\begin{array}{ccc}
 i \sigma _2 & 0 & 0 \\
 0 & -i \sigma _2  & 0 \\
 0 & 0 & -i \sigma _2  \\
\end{array}
\right)\,.
\eeq
The elementary $(t, b)$ are embedded in the spurions  even under the DM parity. We can  write down the options for the left-handed fields:
\beq
D_{L, A}^{1} = \left(
\begin{array}{cccc|cc}
 0 & 0 & 0 & \frac{t_{L}}{\sqrt{2}} & \phantom{0} & \phantom{0} \\
 0 & 0 & 0 & \frac{b_{L}}{\sqrt{2}} & &  \\
 - \frac{b_{L}}{\sqrt{2}} & \frac{t_{L}}{\sqrt{2}} & 0 & 0 &  &  \\ 
0  & 0  & 0 & 0  &  &  \\ \hline
  & &  &  &  &   \\
  &   &   &   &  &   \\
\end{array}
\right) \,, \quad 
D_{L, S}^{2} = \left(
\begin{array}{cccc|cc}
 0 & 0 & 0 & \frac{t_{L}}{\sqrt{2}} & \phantom{0} & \phantom{0}\\
 0 & 0 & 0 & \frac{b_{L}}{\sqrt{2}} & &  \\
  \frac{b_{L}}{\sqrt{2}} & - \frac{t_{L}}{\sqrt{2}} & 0 & 0 &  &  \\ 
0  & 0  & 0 & 0  &  &  \\ \hline
  & &  &  &  &   \\
  &   &   &   &  &   \\
\end{array}
\right)\,,
\eeq
and for the right-handed top:
\beq
D_{R, A}^{1} = \frac{t_R}{2}\begin{pmatrix}
\mathbbm{1}_2 & 0 & 0\\
0 & -\mathbbm{1}_2 & 0\\
0 & 0 & 0
\end{pmatrix}  \,,
~~
D_{R, A}^{2}= \frac{t_R}{2\sqrt{3}}\begin{pmatrix}
\mathbbm{1}_2 & 0 & 0\\
0 & \mathbbm{1}_2 & 0\\
0 & 0 & -2 \mathbbm{1}_2
\end{pmatrix}  \,,
~~
D_{R, S}^{3} = \frac{t_R}{\sqrt{2}} \left(
\begin{array}{ccc}
0 & 0 & 0 \\
0  & \sigma_3 & 0 \\
0 & 0 & 0
\end{array}
\right) \,.
\eeq
The general spurions  are  the linear combinations:
 \begin{eqnarray}
D_{L} &=& Q_{A1} D_{L,A}^{1} + Q_{S2} D_{L,S}^{2} \,,\\
D_{R} &=& R_{A1} D_{R, A}^{1} + R_{A2} D_{R,A}^{2}  + R_{S3} D_{R,S}^{3}  \,.
\end{eqnarray}
Due to the DM parity, there is no tadpole for $H^0, A^0$ and $\eta_{3,4}$. 
To avoid the tadpole for $\eta_1$, we need to impose the condition: 
\beq
Q_{S2}^* Q_{A1}  -  Q_{S2} Q_{A1}^* =0\,.
\eeq
Hence  expanding Eq.~\eqref{eq:Lmix} for the $SU(6)/Sp(6)$ model, we obtain:
\beq \label{eq:LmixSp}
\mathcal{L}_{mix} &=& \frac{1}{2} y_L f \bar{t}_L \left(\sqrt{2} Q_{S2} T_1 \sin (\alpha ) -Q_{A1} \left(\cos (\alpha )
   \left(T-T_X\right)+T_X+T\right)\right)   \nonumber \\    
&-& \frac{1}{2} y_R f \bar{t}_R \left(2 R_{A1} T_1 \cos (\alpha ) +2 R_{A2} T_2 +\sin (\alpha ) R_{S3}  \left(T-T_X\right)\right) \nonumber \\
&-& y_L f \cos \alpha \bar b_L Q_{A1}  + \dots
\eeq
And  the top quark mass is derived to be:
\beq
m_t =  \frac{f^2 M \sin \alpha \left(\sqrt{2} Q_{S2} R_{A1}-Q_{A1} R_{S3}\right) y_L y_R}{2
   \sqrt{M^2+f^2 Q_{A1}^2 y_L^2} \sqrt{M^2+f^2 ( R_{A1}^2 +  R_{A2}^2 ) y_R^2}}
\eeq

\section{Electroweak precision observables} \label{sec:EWPO}

Before discussing the DM property, we investigated the impact of the EW precision observables (EWPO) on the parameter space of the models.  The leading effects can be encoded in the Peskin-Takeuchi parameters~\cite{Peskin:1990zt, Peskin:1991sw}, in particular the $S$ and $T$ parameters, which are typically modified in all composite Higgs models by various sources.

Firstly, the vacuum misalignment via the reduced Higgs coupling to $W, Z$ gauge bosons  leads to the well-known logarithmic contribution:
\beq
\Delta T_h &=& -\frac{3 }{8 \pi \cos^2 \theta_W} \left( \sin^2 \alpha \log \frac{\Lambda }{m_h} + \log \frac{m_h }{m_{h,ref}}\right) \,, \\
\Delta S_h &=& \frac{1}{6 \pi} \left( \sin^2 \alpha \log \frac{\Lambda}{m_h} +  \log \frac{m_h }{m_{h,ref}}\right) \,, \label{TS0}
\eeq
with $\Lambda = 4 \pi f $. The EW  triplets and inert doublet also contribute to $S$ and $T$ at one loop level. The $S$ parameter from the scalars is relatively small, while  the $T$ parameter can be relatively large if there is a large mass splitting in the component fields.   Both the SU(6)/SO(6) and SU(6)/Sp(6) models feature an inert Higgs doublet, whose contribution to $T$ can be estimated as~\cite{Barbieri:2006dq}:
\beq
\Delta T_H &=& \frac{\delta m^2}{24 \pi \sin^2 \theta_W M_W^2} \,, \quad  \delta m^2 =  \left( m_{H^{\pm}} -m_A \right)  \left( m_{H^{\pm}} -m_{H^0}\right)\,.
\eeq
In these  CHMs,  a characteristic property is that the mass splitting  inside each $SU(2)_L$ multiplet is of the order of  a few GeV.  For example,  $\delta m^2 \sim 10 ~\mbox{GeV}^2$  leads to $\Delta T_H \sim 9.0 \times 10^{-5} $, that is negligible compared with the $\Delta T_h $ in the case of  $\sin \alpha \sim 0.1$. This also applies to the  contribution from the triplet pNGB.   

Another important source to EWPO comes from the top sector.  First of all, all the top partners that are odd under the DM parity will not contribute to $S$ and $T$.  As pointed out in~\cite{Cai:2022zqu}, two effects in the top sector are relevant: One stems from the rotation to the mass eigenstates, and it is usually included in the literature; The other emerges from the misalignment effect inherited in the CCWZ objects $d_\mu$ and $E_\mu$, as first considered in Ref.~\cite{Cai:2022zqu}.   Both effects arise at the $\sin^2 \alpha$ order.  In particular, the misalignment violates  the unitarity   and will mix the top partners  in two different multiplets. Following~\cite{Cai:2022zqu},  we can simply  consider a scenario where  a single multiplet of top partners is involved in the partial compositeness mixing. Hence, the contribution from the basis rotation decouples from the misalignment one at the $\mathcal{O}(\sin^2 \alpha)$, so that two effects can be separately computed.

We investigated  two mixing patterns  for the top quark mass generation via  partial compositeness, i.e. the bi-doublet scenario common to both SU(6) models, and  the triplet   $(3,1)+(1,3)$  scenario in SU(6)/SO(6). For the latter, a large negative contribution $\Delta T \lesssim  - 0. 18 $ from the basis rotation  pushes the top partner masses to $\mathcal{O}(10)$ TeV  for small mass splitting in the inert doublet. Henceforth, we relegated the  full results of this  scenario to the Appendix {\ref{Appendix3}}.  Interestingly,  the bi-doublet scenario permits masses of a few TeV order. We will first discuss  the rotation effect in the bi-doublet scenario.  At the zeroth order, the mixing from Eqs~\eqref{eq:LmixSO} and~\eqref{eq:LmixSp} gives: 
\beq
 m_T^{(0)} = m_B^{(0)} = \frac{M}{\cos \phi_L}  
\eeq
with the mass of other top partners to be  $M$ at the leading order. As  the custodial symmetry is conserved up to $\mathcal{O}(\sin^2 \alpha)$ for the bi-doublet mixing, the $T$ parameter received very small contributions $\Delta T_{\rm mix}  = \mathcal{O} (\sin^4 \alpha)$.  Instead,  the mixing contribution to $S$ from  a bi-doublet  can be written in the universal form as follows:
\beq
\Delta S_{\rm mix} &=& \frac{N_c}{2 \pi } \Big[  \frac{m_t^2}{m_T^2} \Big( \frac{\cos^2\phi_L }{\sin^2\phi_L}  \bar{\psi}_+(y_t, y_T)  +  \frac{4}{\sin^2 2 \phi_L} {\bar \psi}_+(y_{T_X}, y_t)  - \frac{\cos^2\phi_L }{\sin^2 \phi_L}  \chi_+(y_t, y_T) \nonumber \\ & - &  
 \frac{4}{\sin^2 2 \phi_L}  \chi_+(y_{T_X}, y_t) \Big) + 2  \bar{\psi}_+(y_T, y_B) +  2 \bar{\psi}_+ (y_X, y_{T_X})  \Big]   \,, \label{SBd}
\eeq
where  the variable is  $y_q = \frac{m_q^2}{m_Z^2}$   and  the explicit expressions of $\chi_\pm$  along with other well-known functions $\psi_\pm, \theta_\pm$  are presented in Appendix~\ref{Appendix3}. The  $\bar \psi_+ $ is a new function defined in~\cite{Cai:2022zqu}:
\beq
&&\bar{\psi }_+ (y_{i}, y_{j}) = \frac{2}{3}\left( Y_{L}^i - Y_{L}^j \right) - \frac{2}{3} Y^{vq} \log \left(\frac{y_i}{y_j}\right) \,, \label{Psibar+}
\eeq
with  $Y^{vq}$ being  the hyper-charge of the vector-like quark. In Eq.~\eqref{SBd}, the coefficients of the first four terms are proportional to $\frac{m_t^2}{m_T^2} \sim \sin^2 \alpha$  that comes from the rotation of  gauge couplings, while the remaining two terms encode the contribution from the mass splitting within one  representation. 
For the  SU(6)/SO(6) coset,  we find that: 
\beq
m_{T} -m_{B} =  \frac{\cos^2 \phi_L}{2 \sin^2 \phi_L} \frac{m_t^2}{m_{T}^{(0)}} - \frac{3 \sin^2 \alpha  \sin^2 \phi_L}{4}  m_{T}^{(0)}  \,, ~~~
 m_{T_X}-m_X = \frac{2\cos \phi_L}{ \sin^2 2 \phi_L} \frac{m_t^2}{m_{T}^{(0)}} \,. 
\eeq
Instead, for the SU(6)/Sp(6) coset,  only the mass difference  among  $(T, B)$ changes:
\beq
m_{T} -m_{B} =  \frac{\cos^2 \phi_L}{2 \sin^2 \phi_L} \frac{m_t^2}{m_{T}^{(0)}} + \frac{\sin^2 \alpha  \sin^2 \phi_L}{4}  m_{T}^{(0)}  \,, ~~~
 m_{T_X}-m_X = \frac{2\cos \phi_L}{ \sin^2 2 \phi_L} \frac{m_t^2}{m_{T}^{(0)}} \,. 
\eeq 

The misalignment  effect will modify the gauge couplings as well. In contrast to the rotation effect,  the  $S, T$ parameters from  the misalignment receive divergent contributions in the low energy effective theory, as the consequence of unitarity violation. In the following we will focus on this effect  in the bi-doublet mixing scenario.

\subsection{SU(6)/SO(6)}

 In the  $SU(6)/SO(6)$ CHM, the  misalignment is encoded in Eq.(\ref{eq:ESO}-\ref{eq:dSO}). Note that the $d_\mu$ term will generate gauge interactions  connecting  the bi-doublet with  the triplets. Substituting  the zeroth order rotation of $(t, T)$ and $(b, B)$  into Eq.~\eqref{eq:Lcomposite}, the custodial symmetry  is broken at $\mathcal{O}(\sin^2 \alpha)$, and we can derive the misalignment contribution to $T$:
\beq
\Delta T_{\rm mis} &=& \frac{N_c \sin^2 \phi_L }{16 \pi \sin^2 \theta_W \cos^2 \theta_W} \Big[ \frac{\kappa^{\prime 2}}{8}\sin^2 \alpha \left[ \theta_+ (y_t, y_M)  - \theta_+ (y_{M}, y_b) \right]  -2 \sin^2 \frac{\alpha}{2}  \theta_+ (y_t, y_b)  \nonumber \\ 
&+&  \left( 2 \sin^2 \frac{\alpha}{2} + \frac{\kappa^{\prime 2}}{8} \sin^2 \alpha \right)  \Big[ (y_t -y_b)  \left( \log\frac{\Lambda^2}{m^2_Z} -\frac{1}{2} \right) -  2 (y_t \log y_t  - y_b \log y_b ) \Big]  \,,
\eeq
where  the second line is related to the divergence $(y_t -y_b) \left( \log\frac{\Lambda^2}{m^2_Z} -\frac{1}{2} \right) $.   Differently from the bi-doublet case of the minimal SU(4)/Sp(4) coset,   the coefficient of  divergent term  is  definite positive  for  $\sin \alpha \neq 0$.  Using the same approach from Ref.~\cite{Cai:2022zqu},  the $\Delta S_{mis}$   turns out to be:
\beq
\Delta S_{mis} &=&   \Delta S_{div} - \frac{N_c }{2 \pi } \frac{\kappa^{\prime 2}}{8}   \sin^2\alpha  \Big[    \sin^2 \phi_L  \left( \chi_+ (y_t, y_M) + 2  \chi_+  (y_b, y_M) \right)   \nonumber \\
& + &3  \left( \cos^2 \phi_R + 1 \right)   \chi_+ (y_{T}, y_M ) + 6 \cos \phi_R \left[  \chi_- (y_T, y_M ) - \psi_- (y_T, y_M ) \right] \Big] \,,
\eeq
with $y_M = \frac{M^2}{m_Z^2}$ standing for any top partner other than $(T, B)$ and  $ \Delta S_{div}$ including all logarithmic terms:
\beq
 \Delta S_{div} &=& \frac{N_c }{2 \pi } \Big[  \sin^2 \phi_L  \Big[  \left( \frac{2}{3} \sin^2 \frac{\alpha}{2} - \frac{\kappa^{\prime 2}}{16} \sin^2 \alpha \right) \left(\frac{1}{3} - \frac{1}{3} \log y_t^2 \right)  -\frac{2}{3} \sin^2  \frac{\alpha}{2}  \Big]  \nonumber \\
& + & (\cos^2 \phi_L  + 1 ) \left( \frac{2}{3} \sin^2 \frac{\alpha}{2} - \frac{3 \kappa^{\prime 2}}{16} \sin^2 \alpha \right)  \left(\frac{1}{3} - \frac{1}{3} \log y_T^2 \right)   \nonumber \\
& + & 2 \left( \frac{34}{3} \sin^2 \frac{\alpha}{2} - \frac{9}{16} \kappa^{\prime 2} \sin^2 \alpha \right) \left(\frac{1}{3} - \frac{1}{3} \log y_M^2 \right) \nonumber \\ &+&  \frac{\kappa^2}{4} \sin^2 \alpha  - \frac{\kappa^{\prime 2}}{8}\sin^2 \alpha \sin^2 \phi_L \left(\frac{1}{3} - \frac{1}{3} \log y_b^2 \right) \nonumber \\ &+& \left( 16 \sin^2 \frac{\alpha}{2} -  \kappa^{\prime 2} \sin^2 \alpha \right)  \left( \log\frac{\Lambda^2}{m^2_Z} -\frac{7}{6} \right)   \Big]\,.
\eeq
In the limit of $\sin \phi_L  = 0$  ( $y_T = y_M $),  similarly to the minimal SU(4)/Sp(4) case,  the misalignment contribution can be  largely simplified to:
\beq
\Delta S_{mis} &=& \frac{N_c}{2 \pi } \left( 16 \sin^2 \frac{\alpha}{2} -  \kappa^{\prime 2} \sin^2 \alpha \right)  \left( \log\frac{\Lambda^2}{m^2_T} -\frac{2}{3} \right)  \label{dS1}
\eeq 
that vanishes  for $\kappa' \simeq 2/  \cos \frac{\alpha}{ 2}$.

\begin{figure}[t]
	\centering 	
	\includegraphics[scale=0.56]{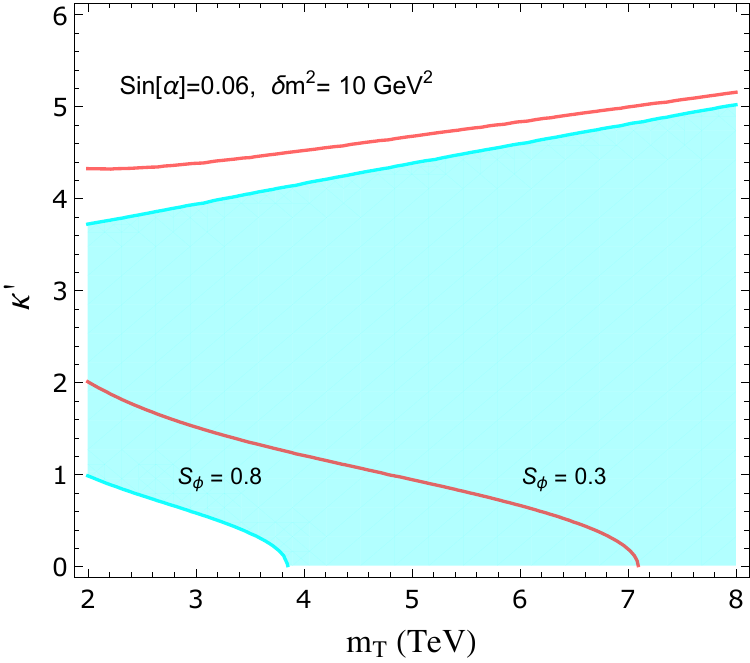}
	\qquad
	\includegraphics[scale=0.56]{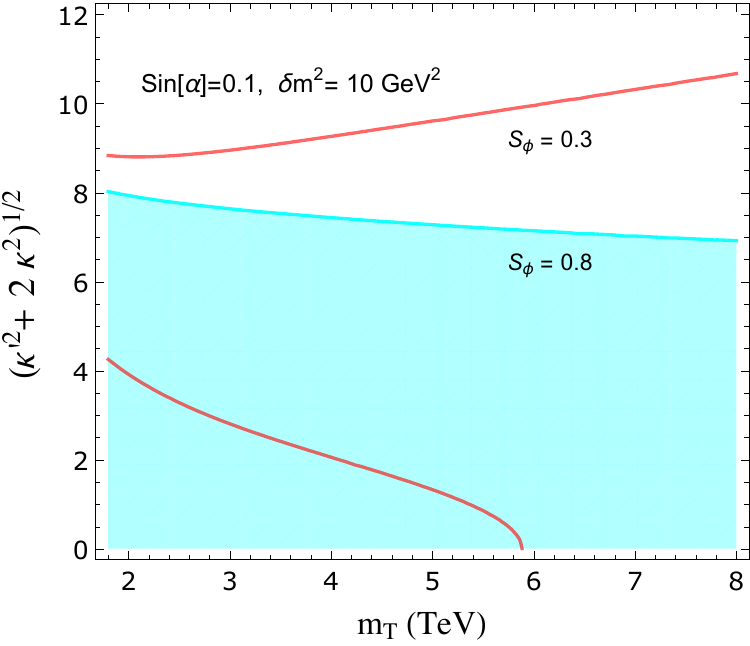}
	\caption{Regions satisfying the  $S,T$  precision constraints  at  $99\%$ C.L. for the bi-doublet mixing scenario, with the left one for $SU(6)/SO(6)$ and the right one for $SU(6)/Sp(6)$.  The cyan region is for $\sin \phi_L = 0.8$ and the region between the two red lines are for $\sin \phi_L =0.3$.}
	\label{fig:EWPT}
\end{figure} 

\subsection{SU(6)/Sp(6)}
 
As we can see from Eq.(\ref{eq:dSP1}-\ref{eq:dSP2}), the  $\kappa$ and $\kappa'$ terms in the SU(6)/Sp(6) case generate  the same interactions between the components fields in the bi-doublet  and one singlet $T_2$ (in $\psi_A$) or $T_3$ (in $\psi_1$). For simplicity, we  take the degenerate limit  $m_{T_2} = m_{T_3} = M$, which gives:
\beq
\Delta T_{\rm mis} &=& \frac{N_c \sin^2 \phi_L}{16 \pi \sin^2 \theta_W \cos^2 \theta_W} \left[ \frac{\kappa'^2 + 2 \kappa^2}{24}   \sin^2\alpha \Big[ \theta_+(y_M, y_b) - \theta_+ (y_M, y_t)\Big] \right.  \nonumber \\   &- & \left. 2 \sin^2 \frac{\alpha}{2}  \theta_+ (y_t, y_b)  
+ \left(2 \sin^2 \frac{\alpha}{2} -\frac{\kappa'^2 +2 \kappa^2 }{24}  \sin^2 \alpha \right)  \nonumber  \right.  \\
&\times& \left.  \Big[ (y_t -y_b)  \left( \log\frac{\Lambda^2}{m^2_Z} -\frac{1}{2} \right) -  2 (y_t \log y_t  - y_b \log y_b ) \Big] \right]  \,.
\eeq 
This contribution  enjoys  the same structure as in the minimal SU(4)/Sp(4).  For $\Delta S_{\rm mis}$, we obtain:
\beq
\Delta S_{\rm mis} &=& \Delta S_{div} - \frac{N_c \sin^2 \alpha}{2 \pi} \Big[ \frac{\kappa^{\prime 2 } + 2 \kappa^2}{24} \left[ \sin^2 \phi_L \chi_+ (y_t, y_{M}) + (\cos^2 \phi_L +1 ) \chi_+ (y_T, y_M ) \right]  \nonumber \\
&+&  \frac{\kappa^{\prime 2 } +2 \kappa^2}{12} \cos \phi_L  \left[\chi_- (y_T, y_M) -\psi_-(y_T, y_M) \right]
\eeq
with 
\beq
\Delta S_{div} &=& \frac{N_c}{2 \pi} \Big[ \frac{8}{3}(2 \sin^2 \frac{\alpha}{2} -\frac{ \kappa'^2 + 2 \kappa^2}{24} \sin^2 \alpha) \left( \log\frac{\Lambda^2}{m^2_Z} -\frac{7}{6} \right)  + \frac{\kappa^{\prime 2} +2 \kappa^2}{36} \sin^2 \alpha   \nonumber \\
&+&  \sin^2 \phi_L \Big[ \left(\frac{2}{3} \sin^2 \frac{\alpha}{2} - \frac{ \kappa'^2 + 2 \kappa^2}{48} \sin^2 \alpha\right) \left(\frac{1}{3} - \frac{1}{3} \log y_t^2 \right) -\frac{2}{3} \sin^2 \frac{\alpha}{2} \Big] \nonumber \\
&+& \left(\cos^2 \phi_L +1\right)\left(\frac{2}{3} \sin^2 \frac{\alpha}{2} - \frac{ \kappa'^2 + 2 \kappa^2}{48} \sin^2 \alpha \right) \left(\frac{1}{3} - \frac{1}{3} \log y_T^2 \right) \nonumber \\
&+& 2 \left(\frac{10}{3} \sin^2 \frac{\alpha}{2} - \frac{ \kappa'^2 + 2 \kappa^2}{16} \sin^2 \alpha \right) \left(\frac{1}{3} - \frac{1}{3} \log y_M^2 \right)  \Big]\,.
\eeq
In analogy to the SU(6)/SO(6) case,  $\Delta S_{\rm mis}$  can be estimated in the limit of $\sin \phi_L =0$ as:
\beq
\Delta S_{mis} &=& \frac{4 N_c}{3 \pi }   \left(2 \sin^2 \frac{\alpha}{2} -\frac{ \kappa'^2 + 2 \kappa^2}{24} \sin^2 \alpha \right)  \left( \log\frac{\Lambda^2}{m^2_T} -\frac{2}{3} \right) \,, \label{dS2}
\eeq
which is consistent with Eq.~\eqref{dS1} with the coefficient difference  originating from  the  representation of the top partners.  Note  that the divergence of $\Delta T_{\rm mis}$ and $\Delta S_{\rm mis}$  simultaneously vanish at the point $\sqrt{\kappa'^2 + 2 \kappa^2} = 2 \sqrt{3} / \cos \frac{\alpha}{2}$.

\subsection{EWPT bounds}

We performed a  $\chi^2$ analysis   for the parameter space spanned by $(\kappa,  \kappa',  m_T, \sin \phi_L, \sin \alpha)$ in  the two SU(6) models,  using  the EW precision data  provided in  the Particle Data Group (2022)~\cite{Workman:2022ynf}. In the bi-doublet mixing scenario, the  positive contribution  from $\Delta T_{mis}$ can be large enough to  compensate the negative one from $\Delta T_h$.  The  plots in Figure~\ref{fig:EWPT} display the regions  permitted  by the  $S$ and $T$  bounds by assuming $U =0$.  The cyan band for $\sin \phi_L =0.8$ and  the region between the two red lines for $\sin \phi_L = 0.3$ are  allowed by EWPO at $99 \%$ C.L.. In fact, the  patterns in two  plots are  mainly determined  by  $\Delta T_{mis}$ as  the $\kappa'^{2}$ and $(\kappa'^2 + 2 \kappa^2)$ contributions have opposite signs in the two models.   We find that, in order to allow  $\kappa' =0 $  in the region of $m_T \sim $ a few TeV,  a small  symmetry breaking angle  is preferred for the SU(6)/SO(6) model.  Instead, in  the  SU(6)/Sp(6) model,  the $\sin \alpha$  can  be of order $0.1$ and smaller under  the EWPO constraint.  In Figure~\ref{fig:EWPT}, we projected  the  EW bounds either on the $(\kappa', m_T)$ plane (left panel) or  on the $(\sqrt{\kappa'^2 + 2 \kappa^2} , m_T)$ plane (right panel).  Both plots show that with  a larger  $\sin \phi_L \sim 0.8$,  the lower bound for  the top partner  can be relieved.

\section{pNGB Dark Matter} \label{sec:DM}

\begin{table}[t!] 
\begin{center}
  {\tabulinesep=2.5mm 
\begin{tabu}{|c|c|c|c|c|c|c|}
\hline    Model & DM & Partner &  $\tilde y_L$  & $\tilde y_R$ & $\lambda_t$ & $a_V$  \\  \hline 
A & $ \eta_2$ & $\tilde{T}_1 $ & $  i \frac{ m_T}{2 \sqrt 2 f }  \sin 2 \phi_L \sin 2\alpha $ &  $ - i  y_d  \frac{m_t}{v} \frac{1}{\sin \phi_L}$ & $\frac{1}{2}(1+ \cos^2 \phi_L)$ & $ \frac{1}{2} \sin^2 \alpha$ \\    \hline  
 \multirow{2}{*}{B} &  $\frac{\eta_3 + i \eta_4}{\sqrt{2}}$  &  $\tilde{X}_{\frac{2}{3}} $ & $\frac{m_T}{4 \sqrt{2} f}  \sin 2 \phi_L  \sin \alpha $  & $  \frac{\sqrt{2} m_t}{v} \frac{1}{\sin \phi_L} $ & \multirow{2}{*}{$\frac{1}{4}  \cos^2 \phi_L $} & \multirow{2}{*}{$ 3 \sin ^2\left(\frac{\alpha }{2}\right)$} \\
\cline{2-5}  & $ \frac{\eta_3 - i \eta_4}{\sqrt{2}}$  &   $\tilde{Y}_{\frac{2}{3}} $ & $- \frac{m_T}{4 \sqrt{2} f}  \sin 2 \phi_L \sin \alpha $ & $- \frac{ \sqrt{2} m_t}{v} \frac{1}{\sin \phi_L} $ & & \\
\hline  
\end{tabu}}
\end{center} 
\caption{The couplings relevant to DM annihilation in  Model A ($SU(6)/SO(6)$ CHM) and  Model B ($SU(6)/Sp(6)$ CHM) for the bi-doublet mixing scenario, with $y_d =  \frac{4  R_{S2}}{\sqrt{3} R_{S1}-\sqrt{2} R_{S2}}$. In the 3rd row, the  partner is a $Z_2$-odd  heavy quark  that  mediates the interaction with a  scalar DM  and  SM tops.} \,  \label{coupling}
\end{table} 

Due to the existence of a dark parity,  the odd pNGBs $(S, H_2)$  and the top partners $\tilde{T}$  in the composite inert Higgs model constitute the dark sector.  In this paper, we  consider the scenario of $m_{\tilde{T}} \gg m_{H_2} >   m_S$, so that   the  singlet $S$  is the DM candidate.  The  rationale is  the inert odd $H_2$ would most likely be excluded by direct detection bounds, due to  a large Z boson coupling. The parity-odd scalars $H_2$ and $S$ were kept in a thermal equilibrium with the SM particles in the early universe and froze out as the temperature drops to  $T_{f} \sim m_S / 25$. When the mass difference between the  doublet $H_2$ and singlet $S$  is large enough, the co-annihilation effect will be suppressed by $e^{-(m_{H_2} -m_S)/T_f}$. Hence for $m_{H_2} > 1.2 \, m_S$,  it is adequate to consider just  the  singlet annihilation.  The Lagrangian  relevant to the DM phenomenology is:   
\beq
\mathcal{L}  \supset &-& C_{h S} \, h S^\dag S  - a_t \frac{m_t}{v}  \, h \bar t t -  \tilde{y}_{L } \, \bar t_{L} \tilde{T}_{R} S  -  \tilde{y}_{R}  \, \bar t_{R} \tilde{T}_{L} S + \lambda_t \frac{m_t}{f^2}S^\dag S \bar t_L t_R  + {\rm h.c.} \nonumber \\
&+& \frac{ 2 m_W^2 }{v}   \left(  h  \cos \alpha - \frac{a_V}{v}  S^\dag S \right) \left(  W^{+\mu} W^-_{\mu} + \frac{Z^{\mu} Z_{\mu}}{ 2 \cos^2 \theta_W} \right) \nonumber \\
&+& \frac{g_2}{4\cos \theta_W} \left( 1- \cos \alpha \right) Z_\mu S^\dag i \overset\leftrightarrow{\partial^\mu} S + Z_\mu \bar q \gamma^\mu \left(g_{v, q} - g_{a, q} \gamma_5\right) q  \,, \label{eq:Ldm}
\eeq
 with $a_t \simeq 1- \mathcal{O} (\sin^2 \alpha)$,  $g_{v, q} =  \frac{g_2}{2 \cos \theta_W} \left( T_L^3  - 2 Q_q \sin^2 \theta_W\right)$ and $g_{a, q} =  \frac{g_2}{2 \cos \theta_W} T_L^3 $. Note that the DM is a real scalar singlet $S = \eta_2$ in SU(6)/SO(6) and a complex one $ S = \frac{1}{\sqrt{2}}(\eta_{3} +i \eta_4)$ in SU(6)/Sp(6).  The $Z$ coupling to DM only exists for the complex singlet, and it is proportional to $(1- \cos \alpha)$ via the EW misalignment effect. The  $S^\dag S \bar t t$  term  is  connected to the generation of top quark mass and derived by expanding the trace part  in Eq.\eqref{eq:PC} till the  $1/f^2$ order. In the mass basis, the coefficient of this quartic term can be expressed  in the form of $\lambda_t \frac{m_t}{f^2}$, that matches  the structure derived from an EFT approach in~\cite{Cai:2020njb}.  In fact  we  can comprehend this operator  as the effect of  integrating out  the even top partners in $\Psi_A$ that mix with SM tops.  In Table~{\ref{coupling}}, we report the dark Yukawa couplings $\tilde{y}_{L,R}$ from the bi-doublet mixing scenario and  the  quartic couplings  $(\lambda_t,  a_V)$ in the two models. For  the right-handed top spurion in SU(6)/SO(6), we set  $R_{S1} = 0, R_{S2} =1$ for a  $ \tilde{y}_{R} $ of $\mathcal{O}(1)$.  Since $m_t$ is generated collectively by the  left and right handed pre-Yukawa spurions,  two  couplings  $\tilde{y}_{L,R}$ are  present in the  Lagrangian to couple the DM with the dark top partners.  This feature extends the simplified vector-like quark portal DM model, where only one chiral (either left or right) Yukawa coupling is taken into account~\cite{Giacchino:2013bta, Cai:2018upp, Colucci:2018vxz}. 

The Higgs portal coupling $C_{hS}$ is generated by the effective potential for the pNGBs and it depends on all the parameters in the underlying theory that break the global symmetries of the strong sector. In principle, the numerical values of this coupling need to  be consistent with the non-tachyon condition (i.e. $m^2_{\rm pNGB} >0$) and with the stability of the dark parity (i.e. the absence of a VEV for the dark pNGBs)~\cite{Cacciapaglia:2019ixa,Cai:2020njb}.  As this exercise is strongly dependent on the details of the model, general and assumption-free bounds cannot be obtained and we will consider $C_{hS}$ as a free positive parameter with a reminder that over large or small values may be unphysical.  Assuming the shift symmetry is dominantly broken  with dark top partners running in the loop, one would expect  $C_{hS}$ to be of  the same order of  the Higgs quartic times its VEV, i.e.  $C_{hS} \sim \lambda_{h^4} v \sim 20$~GeV, as the Higgs mass is generated by the same spurions. However, this value should  not  be too large to conflict with  DM direct detection bounds. Nevertheless,  the DM candidate is not strictly correlated to the Higgs dynamics. With the  cancellation from different  contributions, values of $C_{hS}$  can be  in a few GeV  and  even  smaller in corners of the parameter space,  that agrees with  Ref.~\cite{Balkin:2017aep} for instance.

In this paper we will consider  the DM annihilation $S^\dag S \to  \bar t t, W^{+}  W^-, Z  Z $, that is justified when  the mass region of interest is  $2 m_S < m_t + m_T$. The main focus of this work is on the $\bar t t$ channel, where the top partners directly contribute. The $WW$ and $ZZ$ channels are included as a comparison and to prove that the $tt$ channel dominates. The $hh$ channel is expected to contribute at the same order as $WW$ and $ZZ$ ones, however it receives model-dependent contributions from the pNGB potential. Hence, to keep the focus on the top partners, we do not include it explicitly and consider its effect as an $\mathcal{O} (1)$ uncertainty on the $WW+ZZ$ channels. Note that  the $(\kappa, \kappa')$ terms are neglected as well because they only entail  the derivative couplings of  DM at the leading order. But  non-zero $(\kappa, \kappa')$  can make the parameter space more natural under the EWPO constraint and  have impact on the fractions of  DM annihilation into $\bar t t $ and di-bosons.

\subsection{Relic density computation} 

The DM relic density  is  determined by the thermal average of the annihilation cross section times the velocity $\langle \sigma v \rangle$, expanded  in terms of $v_{rel}$.   We can first calculate  the un-averaged  $\sigma v $  by  summing over all the  kinematically allowed channels:
\begin{align}
& \sigma  v =\sum_{i,j} \frac{k_{ij}} {32 \pi^2  s} \frac{1}{S_{ij}}  \int  d\Omega  |{\cal{M}}_{ij}|^2 \,,  \\
\mbox{with} \phantom{xxx} &  k_{ij} = \left [1-\frac{(m_{i} + m_{j})^2}{s} \right]^{1/2} \left[1-\frac{(m_{i}-m_{j})^2}{s} \right]^{1/2}\,,     
 \end{align}
where  $S_{ij}$ is a symmetric factor  for  identical final states  and $s = E_{tot}^2$ is the Mandelstam variable, with $E_{tot}$ being the total energy in the center of mass frame. Due to  the derivative coupling of $Z$-$S^\dag$-$S$,   its  cross section  is proportional to $(p_{S^\dag} - p_S)^2 = - (s - 4 m_S^2)$. Hence  in the s-wave limit, the Z-portal  interaction does not contribute to relic density. We will focus on the top Yukawa and Higgs mediating channels.  Under the s-wave approximation, the thermal average of   $\langle \sigma v \rangle_{\bar{t} t} $  in the two SU(6) models is: 
\beq
 \langle \sigma v\rangle_{\bar{t} t}  & = & \frac{3 M^2 }{ 4 \pi  \left(M^2+ m_S^2- m_t^2\right)^2}   \left(1- \frac{m_t^2}{m_S^2}\right)^{\frac{3}{2}}  
    \Big[  2  \tilde{y}_L \tilde{y}_R^*   +  \frac{m_t}{M}  \Big[ |\tilde{y}_L|^2+|\tilde{y}_R|^2   \nonumber \\ &+&   \beta_S  \Big( \lambda_t \frac{4 m_S^2 - m_h^2  }{f^2} - \frac{C_{hS} }{v} \Big) \frac{ M^2+m_S^2-m_t^2  }{4 m_S^2 - m_h^2 }  \Big] \Big]^2  \,, \label{eq:ttbar}
 \eeq
 with $\beta_S = 2 $ for a real scalar ($S^\dag = S$) and   $\beta_S = 1$ for a complex scalar, where the difference shows up in the Higgs mediating channel. From Table~{\ref{coupling}}  we can see that $\tilde{y}_L \tilde{y}_R^*$ can be  of  the same order as $\frac{m_t}{M} |\tilde{y}_R|^2 $, while the $\frac{m_t}{M} |\tilde{y}_L|^2 $ term is negligible.  Note that for $C_{hS} =0$ and in the limit of $M \gg m_S \gg m_f $,  our results agree with  the calculation  in Ref.~\cite{Boehm:2003hm}.  In particular,  two dark top partners ($\tilde{X}_{\frac{2}{3}} $, $\tilde{Y}_{\frac{2}{3}} $)  participate in  the annihilation of the complex scalar DM in SU(6)/Sp(6), which is equivalent to the $t$ and $u$ channels for the real scalar DM in SU(6)/SO(6). Thus in our models, there is no $\frac{1}{4} $ factor difference  among the real and complex cases in the  top partner mediating cross section as pointed out by~\cite{Boehm:2003hm}.  
 
Similarly  for the di-boson  annihilation channels, at the leading order, the thermal average of  $\langle  \sigma v\rangle_{VV}$  with $V = W, Z$ is:
\beq
\langle  \sigma v\rangle_{WW} & = & \frac{\pi  \alpha_E^2  \left(4 \frac{m_S^4}{m_W^4}-4 \frac{m_S^2}{m_W^2} +3 \right)\beta_S^2  }{8 m_S^2  s_W^4 \left(4 m_S^2 - m_h^2\right)^2} \sqrt{1- \frac{m_W^2}{m_S^2}} \nonumber \\ 
 &\times& \left( a_V \left(4 m_S^2- m_h^2  \right) -  C_{hS} \, v \cos (\alpha ) \right)^2\,,
 \eeq
 and
 \beq
\langle \sigma v  \rangle_{ZZ}  &= & \frac{\pi  \alpha_E^2 \left(4 \frac{m_S^4}{m_Z^4}-4 \frac{m_S^2}{m_Z^2} +3 \right) \beta_S^2 }{16  m_S^2 s_W^4 c_W^4 \left(4 m_S^2 - m_h^2\right)^2} \sqrt{1- \frac{m_Z^2}{m_S^2}} \nonumber \\
 &\times& \left( a_V \left(4 m_S^2- m_h^2   \right) -  C_{hS} \, v \cos (\alpha ) \right)^2\,.
 \eeq
Without the Higgs portal interaction, the di-boson annihilation rate is roughly proportional to $ m_S^2$   and  becomes larger for heavier DM.   Assuming  $C_{hS}=0$ and $m_S \sim 200$ GeV, we obtain  $\langle  \sigma v \rangle_{WW} +  \langle \sigma v \rangle_{ZZ}  \sim  a_V^2  \beta_S^2 \times  2.0 \times 10^{-23} \mbox{cm}^3/s$. For  a light DM and  $a_V < 10^{-2}$,  the di-boson channel alone can not  saturate the relic density.

From the perspective of the Boltzmann equation, the freeze-out epoch is fixed by the condition that the DM  particles stop to track the equilibrium distribution. By defining $x_f =\frac{m_S}{T_f}$, this freeze-out temperature is determined  as:
 \beq
 x_f  = \log \left[ 0.038 c(c+2) \langle  \sigma v_{\rm tot} \rangle \frac{ g_S m_S M_{\rm PL}}{ \sqrt{g_* x_f}} \right]
 \eeq
 with $c = \sqrt{2}-1$,  $M_{PL} = 1.22 \times 10^{19}$ GeV and $g_* \sim 100$.  Since $g_S$ counts  the internal DM degrees of freedom, we set  $g_S =2 $  for a complex scalar and $g_S =1 $ for a real scalar.  For a total cross section $\langle  \sigma v \rangle_{\rm tot}  \sim 2\times 10^{-26} cm^3/s$ and $m_S \sim 10^2 - 10^3$GeV,  the freeze-out condition gives $x_f \simeq 20-25$.  The relic density can be estimated using the following formula:
 \beq
&\Omega h^2 \approx 
g_S \frac{ 1.07\times 10^9 ~ \rm{GeV}^{-1}}{ \sqrt{g_*(x_f)} M_{\rm PL} \langle  \sigma v_{\rm tot} \rangle / x_f }\,. \label{eq:relic}  
\eeq

\begin{figure}[t]
	\centering 	
	\subfigure[]{\includegraphics[scale=0.66]{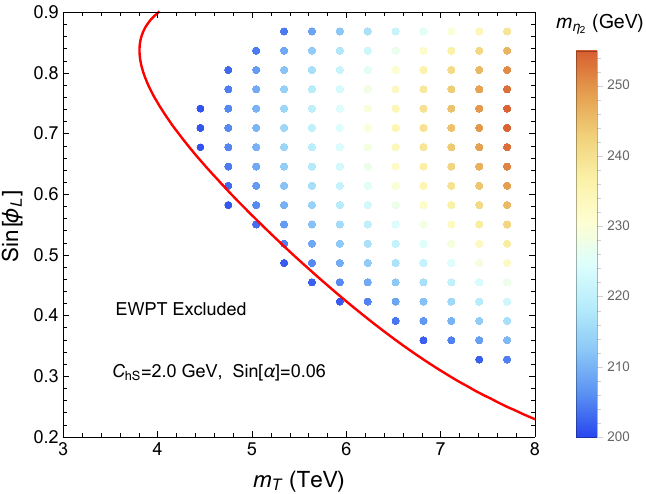}} 
	\quad
	\subfigure[] {\includegraphics[scale=0.66]{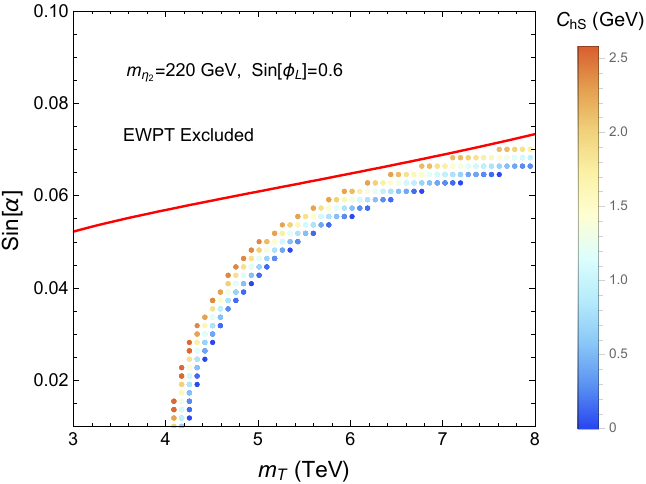}} 
	\subfigure[]{\includegraphics[scale=0.66]{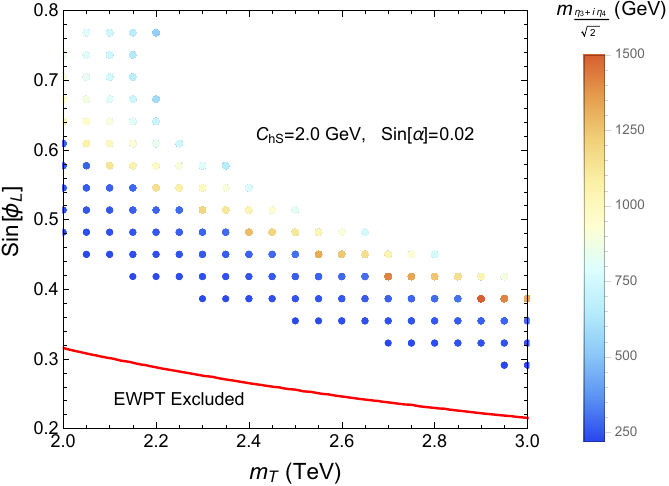}} 
	\quad
	\subfigure[] {\includegraphics[scale=0.66]{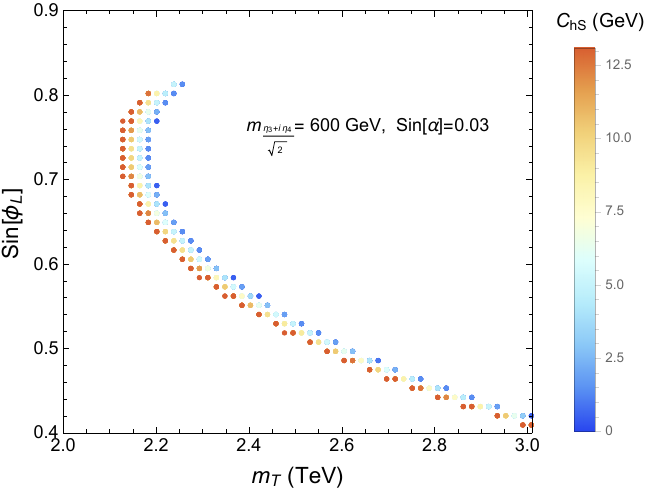}} 
	\caption{The colored regions  provide the correct relic density  $\Omega h^2 = 0.12 \pm 0.002$,  and satisfy the direct detection constraint,  with the upper panels for SU(6)/SO(6) and the lower panels for SU(6)/Sp(6). The red line is the  EWPO bound  at $99 \%$ C.L. with $\kappa = \kappa' =0$.}
	\label{fig:Relic}
\end{figure}

\subsection{Direct detection constraint}

To evaluate the spin-independent  DM-nucleon scattering, we can start with the  effective Lagrangian in terms of  gluons and quarks: 
\beq
\mathcal{L}_{eff} &=& C_q \sum  \limits_{q= u,d,s}   \, m_q S^\dag S  \bar q q  +\frac{\alpha_s}{4 \pi}  \left( 4 \, C_S^g -  C_q - \frac{\lambda_t}{3 f^2}  \right)  S^\dag S G^{A \mu \nu} G^A_{\mu \nu} \nonumber \\
&+& \sum  \limits_{q= u,d} C_Z^q S^\dag i \overset \leftrightarrow{\partial_\mu} S \bar q \gamma^\mu q  \,, \label{Leff}
\eeq
where  the  vector interaction  term is obtained by integrating out the $Z$ gauge boson with
\beq
C_Z^q  = \frac{g_2 \, g_{v,q} }{4  m_Z^2  \cos \theta_W}  \left( 1 - \cos \alpha \right) \,,
\eeq
and  the vector-axial interaction  is not listed as its contribution to the spin-independent cross section is  velocity suppressed. For  the gluon operator in the first line of  Eq.(\ref{Leff}),  the $C_q$ term is generated from  the Higgs portal triangle diagram with heavy quarks $(c, t, b)$ running in the loop, while $C_S^g$ comes from the top partner box diagram.  These  Wilson coefficients in the first line  are given by~\cite{Hisano:2010ct, Cheng:2012qr, Hisano:2015bma}:
\beq
C_q  =  \frac{C_{h S}}{v m_h^2} \,,  \quad  C_S^g  =  -\frac{1}{24 \beta_S} \left[ \frac{\left (\tilde y_L^2 + \tilde y_R^2 \right) m_S^2}{(M^2 - m_S^2 )^2} + \frac{4 \tilde{y }_L \tilde{y}_R M}{(M^2- m_S^2) m_t }\right]\,,
\eeq
where $\beta_S$ takes into account the difference of mediators in two SU(6) models. Then the  coupling $f_{p,n}$ for DM  interacting with  protons or neutrons  are factorized as the Wilson coefficients in Eq.(\ref{Leff}) times the nucleon matrix elements. Expanding in the non-relativistic limit, the spinor structures inside  $ \langle N |  \bar q \gamma^\mu q | N \rangle$, $ \langle N |  \bar q  q | N \rangle$  and $\langle N | G^A_{\mu \nu} G^{A \mu \nu} | N \rangle$ are the same. Hence  the operators from Eq.(\ref{Leff}) contribute to $f_{p,n}$ in an   interference  way:
\beq
f_{p,n} =  \sum \limits_{q= u,d,s}  C_q \, f^{(p,n)}_{q}  +  \frac{2}{9} \,  \left( C_q + \frac{\lambda_t}{3 f^2} -  4 \, C_S^g\right)   f_{TG}^{(p,n)}  \pm \frac{2 m_S}{m_N}  C_Z^{(p, n)}\,, \label{eq:fpn}
\eeq
where the last term is from the $Z$ portal in the  complex scalar DM scenario, with the sign  $\pm$  standing for the scattering of  a particle  or anti-particle DM. Note that $m_N$ is the  nucleon mass and  the form factors  are $C_Z^p =  2 C_Z^u + C_Z^d$,  $C_Z^n =  C_Z^u + 2 C_Z^d$.  For $C_S^g = C_Z^{(p, n)} = 0$, Eq.(\ref{eq:fpn}) goes back to  the pure Higgs portal scenario~\cite{Cai:2020njb}.  And the  matrix elements related to Higgs portal  are defined by~\cite{Drees:1993bu}: 
\beq
f_q^{(p,n)} = \frac{m_q}{m_N} \langle N | \bar q q | N \rangle \,, \quad  f_{TG}^{p,n} = - \frac{9}{8\, m_N}  \langle N | \frac{ \alpha_s}{\pi}G^A_{\mu \nu} G^{A \mu \nu} | N \rangle
\eeq
Operating the nucleon states on the trace of the stress energy tensor  gives the following relation~\cite{Shifman:1978zn}:
\begin{equation}
 f_{TG}^{(p,n)}=  \left( 1- \sum \limits_{q = u,d,s} f_q^{(p,n)}\right)\,.  
\end{equation}
where $f_q^{(p,n)}$ is the quark mass fraction that can be calculated  by lattice simulation  or  chiral perturbation theory, and we will use the values listed in \cite{Hisano:2015bma}.  Finally, the spin-independent cross section is 
\begin{eqnarray}
\sigma_{\rm SI} =  \frac{m_N^2 \beta_S^2 }{ 4 \pi m_S^2} \left( \frac{m_N m_S}{m_N + m_S} \right)^2  \frac{(Z f_p + (A-Z) f_n  )^2}{A^2}\,. \label{eq:SI}
\end{eqnarray} 
As indicated in Eq.(\ref{eq:fpn}),  the Z-portal interaction of  particle and anti-particle DMs is either  constructive or deconstructive with the Higgs portal one in the DM-nucleon coupling $f_{p, n}$.  For  the complex DM in SU(6)/Sp(6) CHM, we  need to average over this two effects, i.e. $\sigma_{\rm SI}^{tot} = \frac{1}{2} \left( \sigma_{\rm SI} +\bar \sigma_{\rm SI} \right)$. The measurement of $\sigma_{\rm SI}$ at the underground detectors~\cite{XENON:2018voc, Cui:2017nnn, Akerib:2016vxi, PandaX-4T:2021bab, LZ:2022ufs} will impose constraint on $(m_S, f_{p,n})$. With $m_T > 2 $ TeV and $m_S  \in (0.2, 1.5)$ TeV,  the  dark top partner induced $\sigma_{\rm SI}$  normally  is $< 10^{-48} \, \mbox{cm}^2$,  several orders of magnitude smaller than the upper limit.  Hence, the direct detection mainly  constrains  the parameters related to the Higgs or Z-portal interactions, i.e.   $(C_{hS}, m_S)$ in SU(6)/SO(6) CHM and  $(C_{hS}, m_S, \sin \alpha)$ in SU(6)/Sp(6) CHM.  For the complex DM,  the Z-portal interaction places a stringent upper bound on $\sin \alpha$. Using the latest LUX-ZEPLIN measurement \cite{LZ:2022ufs}, we find  with $C_{hS} =0$ and $m_S \in (0.2, 1.5)$TeV, the upper limit of $\sin \alpha$  in  SU(6)/Sp(6) CHM  falls in the range of $(0.024, 0.04)$. Adding the Higgs coupling $C_{hS}$ will further reduce the allowed $\sin \alpha $ value.  The benefit to include odd top partners is that the dark Yukawa interaction  brings in sufficient additional  annihilation rate that compensates the smallness of Higgs portal one, while  is not  sensitive to the direct detection in the preferred $(m_T, m_S)$ parameter space.  Therefore,  the dark top partners in fact relieve the tension in a  Higgs portal DM model.

\subsection{Combined analysis} 
  
There are  5  independent parameters in our DM models, i.e. $(C_{hS}, m_S, m_T, \sin \alpha, \sin \phi_L)$, where the last 3 ones are related to EWPOs. A comprehensive analysis is conducted to find the viable parameter space. However  we will not include the bound from indirect detection, e.g. the gamma rays observed in dwarf spheroidal galaxies or the Galactic Center. The signals normally  impose constraint on the low DM mass scenario ($\lesssim100$ GeV)~\cite{Ackermann:2015zua, Hoof:2018hyn}. A recent interpolation  pushed up the bounds  near several hundred GeV for the $\bar b b $, $WW$, $ZZ$ channels,  but lack  of  $\bar t t$ information \cite{Abazajian:2020tww}.  Note that our analyses in both CHMs  are restricted to  a  small $\sin \alpha$ and this results in  $ S S \to \bar t  t$ playing a dominant role  with more than $90 \%$ contribution to the DM annihilation.  For the  DM mass in $(0.2, 1.5)$ TeV,  the upper limit on  $\langle \sigma v \rangle_{\bar tt} $  is around $5.0\times 10^{-26} \rm{cm}^3 s^{-1}$  from the 11 years of Fermi-LAT observations \cite{Hoof:2018hyn},  with almost no  constraint on the parameter space.  

In Figure~{\ref{fig:Relic}},  we  require the scalar DM to accommodate   the correct  relic density  $ 0.118 < \Omega h^2 < 0.122 $~\cite{Aghanim:2018eyx} and simultaneously satisfy the constraints from the direct detection  and EWPOs.  In addition,  collider experiments impose direct constraints on the  dark top partners and the $(T_X, X)$ in the $Z_2$-even bi-doublet~\cite{ATLAS:2018ziw, ATLAS:2018alq, CMS:2018zkf}.   Henceforth,  the points in the plots  need to satisfy $M = m_T \cos \phi_L > 1.3$ TeV.  
The upper panels  refer to SU(6)/SO(6): In Figure~{\ref{fig:Relic}}(a), we set  $C_{hS} =2 $ GeV, and the relic density  dominantly comes from the  top partner channel.  We can  see that  the DM mass is constrained to be $ 200 \lesssim m_{\eta_2} \lesssim 255 $ GeV for $m_T \leq 8 $ TeV.   A lower  DM  mass  from this region will demand for smaller $\sin \phi_L$ and $m_T$, thus  is excluded by EWPOs.  Corresponding results for SU(6)/Sp(6) are displayed in the lower panels.  For this case, the allowed value of $\sin \alpha$ derived from  the direct detection strictly constrains the lower limit of DM mass, while the bound from EWPO is relatively loose. In Figure~\ref{fig:Relic}(c)  with $C_{hS} =2 $ GeV, $\sin \alpha =0.02$,  the DM mass populates in a wide range  $ 200 \lesssim m_{\frac{\eta_3 +i \eta_4}{\sqrt{2}}}\lesssim 1500 $ GeV.  Increasing $\sin \alpha$ will  cut off the lower mass region.  The impacts of $C_{hS}$  for two models  are  visualized in Figure~{\ref{fig:Relic}}(b) and {\ref{fig:Relic}}(d), with  its value  limited by the direct detection bound. From Eq.\eqref{eq:ttbar}, we find  that a positive $C_{hS}$ makes the Higgs portal deconstructive with the  top partner channel, hence requires a lighter  top partner.  The narrow bands from the variation of  $C_{hS}$  indicate  that  the Higgs portal only offers a subleading contribution.

\section{Conclusion} \label{sec:6}  

Extending the global symmetries of composite Higgs models allow to accommodate for stable pNGBs, which can play the role of Dark Matter. In fundamental CHMs with real and pseudo-real realizations, this amounts to add  the HC fermions  that are odd under a   $\mathbb{Z}_2$ dark parity. In return, odd resonances as condensations of  HC fermions appear in all sectors of the theory. However for the complex realization of CHM,   a simple $\mathbb{Z}_2$ for the HC fermions is not  adequate  as  the  conjugate operation plus  internal flavor rotation might be involved.

Partial compositeness for the top quark mass generation plays a crucial role in this class of models. On one hand, the couplings of the elementary top fields must preserve the dark parity, hence imposing non-trivial constraints on the UV completions of the models.  On the other hand, the presence of odd top partners relieves the tension of Higgs portal coupling with the direct detection.

We have considered in detail the role played by  the dark top partners in two models based on fundamental dynamics, where the composite Higgs boson and Dark Matter stem from the cosets SU(6)/SO(6) and SU(6)/Sp(6). Following Ref.~\cite{Cai:2022zqu}, we considered the impact of electroweak constraints on the models. Furthermore, we studied the properties of the pNGB DM  in the presence of the dark top partners. They enter as mediators for the annihilation of the DM  candidate in the early Universe and at direct detection experiments. This effect can dominate over the Higgs mediated processes, with non-trivial interference effects taken into account. As a result, phenomenological constraints require the mass of the even top partner to be in the multi TeV regime, with viable masses above 4 TeV for SU(6)/SO(6) and lower to 2 TeV for SU(6)/Sp(6). These masses can only be directly probed at future high-energy hadronic colliders.

\section*{Acknowledgments}
H.C. \ is supported by the National Research Foundation of Korea (NRF) grant funded by the Korea government (MEST) (No. NRF-2021R1A2C1005615).

\newpage

\appendix  

\section{EW gauge interaction}\label{Appendix1}

We  will  write  down the EW gauge interactions with  the top partners not rotated into  the mass basis. In the $SU(6)/SO(6)$ CHM,  the $\Psi_A$ gauge interaction includes a standard part:
\beq
&& Tr \left[\bar \psi_A \gamma^\mu \left(V_\mu \psi_A + \psi_A V_\mu^T \right) + g_1 B_\mu^0 \hat X \bar \psi_A \gamma^\mu \psi_A  \right] \nonumber \\
&=& \frac{g_2 W^+_\mu}{\sqrt{2}}  \left[ \overline{\tilde{T}} \gamma^\mu \tilde{B} +\overline{\tilde{X}}  \gamma^\mu \tilde{T}_X 
+ \bar{T} \gamma^\mu B +\bar{X}  \gamma^\mu T_X+\sqrt{2}
   \left(  \bar{X}_{\frac{2}{3}} \gamma^\mu X_{-\frac{1}{3}}- \bar{X}_{\frac{5}{3}}  \gamma^\mu X_{\frac{2}{3}} \right) \right] + h.c. \nonumber \\
 &+& \frac{g_2 W^3_\mu}{2}   \Big[ \overline{\tilde{T}} \gamma^\mu \tilde{T} -  \overline{\tilde{B}} \gamma^\mu  \tilde{B} +  \overline{\tilde{X}} \gamma^\mu \tilde{X}  - \overline{\tilde{T}}_X \gamma^\mu \tilde{T}_X + \bar{T} \gamma^\mu T  -  \bar{B} \gamma^\mu B +  \bar{X} \gamma^\mu  X  -  \bar{T}_X \gamma^\mu T_X  \nonumber \\ & -  & 2  \bar{X}_{-\frac{1}{3}} \gamma^\mu X_{-\frac{1}{3}} + 2 \bar{X}_{\frac{5}{3}} \gamma^\mu X_{\frac{5}{3}} \Big] +  g_1 B_\mu^0  \Big[ \frac{1}{6} \left( \overline{\tilde{T}} \gamma^\mu \tilde{T} +  \overline{\tilde{B}} \gamma^\mu  \tilde{B} + \bar{T} \gamma^\mu T +  \bar{B} \gamma^\mu B  \right)  \nonumber \\ &+ &  \frac{7}{6}  \left( \overline{\tilde{X}} \gamma^\mu \tilde{X}  + \overline{\tilde{T}}_X \gamma^\mu \tilde{T}_X +  \bar{X} \gamma^\mu  X +  \bar{T}_X \gamma^\mu T_X  \right) + \frac{2}{3} \left( \bar{X}_{-\frac{1}{3}} \gamma^\mu X_{-\frac{1}{3}} +  \bar{X}_{\frac{2}{3}} \gamma^\mu X_{\frac{2}{3}}  +  \bar{X}_{\frac{5}{3}} \gamma^\mu X_{\frac{5}{3}} \right) 
 \nonumber \\ &-& \frac{1}{3} \bar{Y}_{-\frac{1}{3}} \gamma^\mu Y_{-\frac{1}{3}} +  \frac{2}{3} \bar{Y}_{\frac{2}{3}} \gamma^\mu Y_{\frac{2}{3}} + \frac{5}{3} \bar{Y}_{\frac{5}{3}} \gamma^\mu Y_{\frac{5}{3}}  \Big] 
\eeq
with $V_\mu = g_2 W_\mu^i T^i_L + g_1 B_\mu^0  T^3_R$.  And  the misaligned effect  is separately  encoded in  $\delta E_\mu = E_\mu -V_\mu$.  We can expand the relevant Lagrangian  to obtain at the leading order: 
\beq
&&  Tr \left[ \bar  \psi_A \gamma^\mu  \left( \delta E_\mu \psi_A + \psi_A  \delta E_\mu^T \right) \right]  \supset  \nonumber \\
& & - \frac{g_2  \sin ^2\left(\frac{\alpha }{2}\right)}{2}  \Big[ W_{\mu }^1 \Big[
   \left( \overline{\tilde{B}} - \overline{\tilde{X}}  \right) \gamma ^{\mu }
   \left(\tilde{T}-\tilde{T}_X\right) +  \left( \bar{B} - \bar{X} \right) \gamma ^{\mu } \left(T-T_X\right)  \nonumber \\ & & -\sqrt{2} \left(
   \bar{X}_{\frac{2}{3}} \gamma ^{\mu } \left(X_{\frac{5}{3}}-X_{-\frac{1}{3}}\right) +
   \bar{Y}_{\frac{2}{3}} \gamma ^{\mu } \left(Y_{-\frac{1}{3}}-Y_{\frac{5}{3}}\right) \right)\Big]
\nonumber \\
&& +  i  W_{\mu }^2
   \Big[  \left(\overline{\tilde{B}}+\overline{\tilde{X}}\right) \gamma^\mu \left(\tilde{T}-\tilde{T}_X\right) +
   \left(\bar{B}+\bar{X}\right) \gamma^\mu \left(T-T_X\right) \nonumber \\ & & -  \sqrt{2} \left(
   \bar{X}_{\frac{2}{3}} \gamma^\mu \left(X_{-\frac{1}{3}}+X_{\frac{5}{3}}\right) -
   \bar{Y}_{\frac{2}{3}} \gamma^\mu \left(Y_{-\frac{1}{3}}+Y_{\frac{5}{3}}\right) \right) \Big]
\nonumber \\
&+& 2  \Big( W_{\mu }^3 - B_{\mu }^0 \tan (\text{$\theta $w}) \Big) \Big[  \overline{\tilde{T}} \gamma^\mu \tilde{T}
  - \overline{\tilde{T}}_X \gamma^\mu \tilde{T}_X +  \bar{T} \gamma^\mu T - \bar{T}_X  \gamma^\mu T_X 
    \nonumber \\ && -   \bar{X}_{-\frac{1}{3}} \gamma^\mu X_{-\frac{1}{3}} +
   \bar{X}_{\frac{5}{3}} \gamma^\mu X_{\frac{5}{3}} +  \bar{Y}_{-\frac{1}{3}} \gamma^\mu Y_{-\frac{1}{3}} -
   \bar{Y}_{\frac{5}{3}} \gamma^\mu Y_{\frac{5}{3}} \Big] \Big] \,, \label{eq:ESO}
\eeq
The $\kappa'$ term in Eq.(\ref{eq:Lcomposite}) will also modify the gauge interactions: 
\beq
 && Tr[ \bar \psi_A  d_\mu \gamma^\mu \psi_A] \supset \nonumber \\
 & & \frac{1}{8}  g_2 W_1^\mu  \sin (\alpha ) \left(\sqrt{2} 
   \left( \overline{\tilde{B}}-\overline{\tilde{X}}\right) \gamma_\mu \tilde{T}_1 -  \left(\bar{B}+\bar{X}\right) \gamma_\mu \left(X_{\frac{2}{3}}+Y_{\frac{2}{3}}\right)
\nonumber \right. \\ &+ & \left. \sqrt{2} \bar{T} \gamma_\mu \left(X_{\frac{5}{3}}-Y_{-\frac{1}{3}}\right) + \sqrt{2} \bar{T}_X \gamma_\mu
   \left(Y_{\frac{5}{3}}-X_{-\frac{1}{3}} \right)\right) \nonumber \\
&+&\frac{i}{8} g_2 W_2^\mu \sin (\alpha ) \left(\sqrt{2}  \left(\overline{\tilde{B}}+\overline{\tilde{X}}\right) \gamma_\mu \tilde{T}_1 
   - \left(\bar{B}-\bar{X}\right) \gamma_\mu  \left(X_{\frac{2}{3}}+Y_{\frac{2}{3}}\right) \right. \nonumber  \\ 
   &+& \left. \sqrt{2} \bar{T} \gamma_\mu \left(X_{\frac{5}{3}}+Y_{-\frac{1}{3}}\right)+\sqrt{2}  \bar{T}_X \gamma_\mu \left(X_{-\frac{1}{3}}+Y_{\frac{5}{3}}\right) \right) \nonumber \\ &+& \frac{1}{8} g_2 \left(W_3^\mu -   B_0^\mu \tan (\text{$\theta $w})  \right) \sin (\alpha ) \left(\sqrt{2}  \left( \overline{\tilde{T}} + \overline{\tilde{T}}_X\right) \gamma_\mu \tilde{T}_1+\sqrt{2} \bar{B} \gamma_\mu \left(X_{-\frac{1}{3}}+Y_{-\frac{1}{3}}\right) \right. \nonumber \\ &+ & \left. \left(\bar{T} -\bar{T}_X \right) \gamma_\mu \left(X_{\frac{2}{3}}-Y_{\frac{2}{3}}\right)  +\sqrt{2} \bar{X} \gamma_\mu \left(X_{\frac{5}{3}}+Y_{\frac{5}{3}}\right)\right)  + h.c. \, \label{eq:dSO}
\eeq
Note that  because the charge operator $Q$ is conserved,  only $W_\mu^\pm$ and $Z_\mu$ couplings are misaligned  by the $\sin \alpha$  corrections.

Similarly for the $SU(6)/Sp(6)$ CHM, the Lagrangian of the covariant gauge interaction can split  into the SM  part plus the misalignment one as following:
\beq
&& Tr \left[\bar \psi_A \gamma^\mu \left(V_\mu \psi_A + \psi_A V_\mu^T \right) + g_1 B_\mu^0 \hat X \bar \psi_A \gamma^\mu \psi_A  \right] \nonumber \\
&=& \frac{g_2 W^+_\mu}{\sqrt{2}} \left[ \overline{\tilde{T}} \gamma^\mu \tilde{B} +\overline{\tilde{X}}  \gamma^\mu \tilde{T}_X  + \bar{T} \gamma^\mu B +\bar{X} \gamma^\mu T_X \right] +h.c.\nonumber\\
&+&  \frac{g_2 W^3_\mu}{2}   \Big[ \overline{\tilde{T}} \gamma^\mu \tilde{T} -  \overline{\tilde{B}} \gamma^\mu  \tilde{B} +  \overline{\tilde{X}} \gamma^\mu \tilde{X}  - \overline{\tilde{T}}_X \gamma^\mu \tilde{T}_X +   \bar{T} \gamma^\mu T - \bar{B} \gamma^\mu B \nonumber \\ &+&  \bar{X} \gamma^\mu  X  -  \bar{T}_X \gamma^\mu T_X \Big] + +  g_1 B_\mu^0  \Big[ \frac{1}{6} \left( \overline{\tilde{T}} \gamma^\mu \tilde{T} +  \overline{\tilde{B}} \gamma^\mu  \tilde{B} + \bar{T} \gamma^\mu T +  \bar{B} \gamma^\mu B  \right)  \nonumber \\ &+ &  \frac{7}{6}  \left( \overline{\tilde{X}} \gamma^\mu \tilde{X}  + \overline{\tilde{T}}_X \gamma^\mu \tilde{T}_X +  \bar{X} \gamma^\mu  X +  \bar{T}_X \gamma^\mu T_X  \right) - \frac{1}{3}  \bar{X}_{-\frac{1}{3}} \gamma^\mu X_{-\frac{1}{3}} +  \frac{5}{3}  \bar{X}_{\frac{5}{3}} \gamma^\mu X_{\frac{5}{3}}   \nonumber \\ &+& \frac{2}{3} \left( \bar{X}_{\frac{2}{3}}  \gamma^\mu X_{\frac{2}{3}}  + \bar{Y}_{\frac{2}{3}} \gamma^\mu Y_{\frac{2}{3}} + \bar{T}_{1}  \gamma^\mu T_{1}  + \bar{T}_{2} \gamma^\mu T_{2}  \right)
\eeq 
\beq
&&   Tr \left[ \bar  \psi_A \gamma^\mu  \left( \delta E_\mu \psi_A + \psi_A  \delta E_\mu^T \right) \right]   \supset  \nonumber \\
& & - \frac{g_2  \sin ^2\left(\frac{\alpha }{2}\right)}{2} \Big[  \left(W_\mu^1- i W_\mu^2 \right) \Big[  \overline{\tilde{T}} \gamma^\mu  \tilde{B} +  \overline{\tilde{X}} \gamma^\mu \tilde{T}_X  + \overline{\tilde{Y}}_{\frac{2}{3}}  \gamma^\mu \tilde{X}_{-\frac{1}{3}} + \overline{\tilde{X}}_{\frac{5}{3}} \gamma^\mu \tilde{X}_{\frac{2}{3}}  \Big] \nonumber \\ &+  & \Big[ W_\mu^1  \left(\bar{B}+\bar{X}\right) \gamma^\mu \left(T_X+T\right) + i W_\mu^2  \left(\bar{B}-\bar{X}\right) \gamma^\mu \left(T_X+T\right) \Big] + h.c.  \nonumber \\
  &+&\Big( W_{\mu }^3 - B_{\mu }^0 \tan (\text{$\theta $w}) \Big)  \Big[ \overline{\tilde{T}} \gamma^\mu \tilde{T}  - \overline{\tilde{B}} \gamma^\mu \tilde{B}  +\overline{\tilde{X}} \gamma^\mu \tilde{X}  -  \overline{\tilde{T}}_X \gamma^\mu \tilde{T}_X + \overline{\tilde{X}}_{-\frac{1}{3}} \gamma^\mu \tilde{X}_{-\frac{1}{3}}  \nonumber \\ &+&  \overline{\tilde{X}}_{\frac{2}{3}} \gamma^\mu \tilde{X}_{\frac{2}{3}}-  \overline{\tilde{X}}_{\frac{5}{3}} \gamma^\mu \tilde{X}_{\frac{5}{3}} -  \overline{\tilde{Y}}_{\frac{2}{3}} \gamma^\mu \tilde{Y}_{\frac{2}{3}} -2  \bar{T}_X \gamma^\mu T_X +2 \bar{T} \gamma^\mu T \Big] \Big] 
\eeq
Finally the $\kappa', \kappa$ terms in Eq.(\ref{eq:Lcomposite}) from the CCWZ formalism can be expanded as:
\beq
& &  Tr \left[ \bar{\psi}_A \not{d}\  \psi_A \right] \supset  \nonumber \\
&  & \frac{g_2 \sin (\alpha ) }{8 \sqrt{3}}  \Big[   \sqrt{3} \left(W_{\mu}^1 - i W_\mu^2 \right) \left(
   \overline{\tilde{Y}}_{\frac{2}{3}} \gamma^\mu \tilde{B} +\overline{\tilde{X}}_{\frac{5}{3}} \gamma^\mu \tilde{T}_X-
   \overline{\tilde{T}} \gamma^\mu \tilde{X}_{-\frac{1}{3}} - \overline{\tilde{X}} \gamma^\mu \tilde{X}_{\frac{2}{3}} \right)  \nonumber \\&+& \sqrt{2} \Big(W_{\mu}^1
   \left(\bar{B}+\bar{X}\right) \gamma^\mu T_2 + i W_\mu^2  \left(\bar{B}-\bar{X}\right) \gamma^\mu T_2 \Big)  \nonumber \\ 
  &+& \left(W_{\mu }^3- B_{\mu } \tan (\text{$\theta $w})\right) \Big(\sqrt{3} \Big(
   \overline{\tilde{B}} \gamma^\mu \tilde{X}_{-\frac{1}{3}} +
   \overline{\tilde{T}}_X \gamma^\mu \tilde{X}_{\frac{2}{3}} \nonumber \\ &+&
   \overline{\tilde{T}} \gamma^\mu \tilde{Y}_{\frac{2}{3}} + \overline{\tilde{X}}_{\frac{5}{3}} \gamma^\mu  \tilde{X}\Big)
   + \sqrt{2}  \left(\bar{T}-\bar{T}_X\right) \gamma^\mu T_2 \Big) + h.c. \, \label{eq:dSP1}
\eeq
\beq
 Tr[ \bar \psi_{A}  d_\mu \gamma^\mu \psi_1] + h.c.    
 &\supset &  \frac{g_2  \sin (\alpha )}{4 \sqrt{3}}  \Big[ W^1_{\mu } \left(\bar{B}+\bar{X}\right)  \gamma^{\mu } T_3  + i W^2_{\mu } \left(\bar{B}-\bar{X}\right)  \gamma^{\mu } T_3  \nonumber \\
 &+&  \left( W^3_{\mu }  - B^0_{\mu }  \tan (\text{$\theta $w})  \right)  \left(\bar{T}-\bar{T}_X\right) \gamma^{\mu }  T_3   \Big] + h.c. \, \label{eq:dSP2}
\eeq

\section{DM Yukawa interaction}\label{Appendix2}

We will derive the couplings between the odd pNGB and dark top partners  generated from  the partial compositeness in Eq.(\ref{eq:PC}). For the $SU(6)/SO(6)$ CHM, we find that:
 \beq
\mathcal{L}_{\eta_2} &=& - i \eta _2 y_L \bar{t}_L  \left(  \sin 2 \alpha  Q_{A1}  \tilde{T}_{1} + \sqrt{2} Q_{S2} \left(\tilde{T} \cos^2 \frac{\alpha}{2} -  \tilde{T}_{X}  \sin^2 \frac{\alpha}{2} \right)\right) \nonumber \\
&+& \frac{i}{2}  \eta _2 y_R \bar{t}_R  \left(\sin \alpha  R_{A4}  \left(\tilde{T}_{X }+ \tilde{T} \right)- \tilde{T}_{1} \left( \sin ^2 \alpha  (\sqrt{6} R_{S1} -2 R_{S2}) + 4 R_{S2}\right)\right)  \nonumber \\
 &- & i   \eta _2   y_L    \bar{b}_L \tilde{B} \sqrt{2} \cos \alpha  Q_{S2} + h.c.
 \eeq
 
 \beq
\mathcal{L}_{H_0/A_0} &=& - \frac{1}{2} H_0 y_L \bar{t}_L \left(\sqrt{2}  Q_{A1} \left( \sin 2 \alpha
  \left(\tilde{T}-\tilde{T}_X\right)+ \sin \alpha  \left(\tilde{T}_X+\tilde{T} \right) \right) + 2  Q_{S2} \tilde{T}_1 \right) \nonumber \\
   &+& \frac{1}{2} H_0 y_R \bar{t}_R \left( \cos ^2 \alpha ( \sqrt{3} R_{S1} - \sqrt{2} R_{S2}) + 2 \sqrt{2}  R_{S2} \right) \left(\tilde{T}-\tilde{T}_X\right) \nonumber \\
  &+& \frac{i}{2} A_0 y_L \bar{t}_L \left(\sqrt{2} \sin \alpha  Q_{A1}
   \left(\tilde{T}-\tilde{T}_X\right)+2  \cos \alpha  Q_{S2} \tilde{T}_1 \right) \nonumber \\
  &-& \frac{i}{2} A_0 y_R \bar{t}_R \left(\sqrt{2}  \sin \alpha 
   R_{A4} \tilde{T}_1 - \left(\sqrt{3} R_{S1}+\sqrt{2} R_{S2}\right) \left(\tilde{T}_X+\tilde{T}\right) \right) \nonumber \\
   &-& \sqrt{2} H_0  y_L \bar{b}_L \tilde{B} \sin \alpha Q_{A1} + h.c.
 \eeq
In the $SU(6)/Sp(6)$ CHM, due to the remaining global $U(1)$ symmetry,  the parity odd PNGB are complex scalars. The relevant couplings are:
\beq
\mathcal{L}_{\eta_3+ i\eta_4} &=& \frac{1}{4}  y_L \bar{t}_L \sin \alpha  Q_{A1} \left(\left(\eta _3+i \eta _4\right)
   \tilde{X}_{\frac{2}{3}}- \left(\eta _3-i \eta _4\right) \tilde{Y}_{\frac{2}{3}}\right) \nonumber \\ &-&  \frac{1}{2}  y_L \bar{t}_L Q_{S2} \left(\left(\eta _3-i
   \eta _4\right)  \tilde{T} \cos^2 \frac{\alpha}{2}  +\left(\eta _3+i \eta _4\right) \tilde{T}_X \sin^2 \frac{\alpha}{2}  \right)
\nonumber \\ 
&+& \frac{1}{2 \sqrt{2}} y_R \bar{t}_R  \sin \alpha  R_{A1} \left( (\eta _3 + i \eta_4)
   \tilde{T}_X  - (\eta_3 - i \eta _4) \tilde{T}\right) \nonumber \\ &-&  \frac{1}{2} y_R \bar{t}_R \cos^2 \frac{\alpha}{2}  R_{S3}  \left(\left(\eta _3+i \eta _4\right) \tilde{X}_{\frac{2}{3}}-\left(\eta _3-i \eta _4\right) \tilde{Y}_{\frac{2}{3}}\right)
\nonumber \\
&-&\frac{1}{2} \left(\eta _3-i \eta _4\right) y_L \bar{b}_L \left(\tilde{B}
   Q_{S2}-\tilde{X}_{-\frac{1}{3}} \sin \alpha  Q_{\text{A1}}\right) + h.c.
\eeq
 
\beq
 \mathcal{L}_{H_0 +i  A_0} &=& \frac{1}{4} \left(H_0-i A_0\right) y_L \bar{t}_L \left(\tilde{T} \sin \alpha
   Q_{A1} + 2 \tilde{Y}_{\frac{2}{3}} \sin^2 \frac{\alpha}{2}  Q_{S2}\right) \nonumber \\
   &+&  \frac{1}{4} \left(H_0+i A_0\right) y_L \bar{t}_L \left(2 \tilde{X}_{\frac{2}{3}} \cos^2\left(\frac{\alpha }{2}\right) Q_{S2}- \tilde{T}_X \sin \alpha  Q_{A1} \right) \nonumber \\ &-& \frac{1}{4}  \left(H_0 - i A_0 \right) y_R \bar{t}_R \left(\sqrt{2} \tilde{Y}_{\frac{2}{3}} \sin \alpha 
   R_{A1}-2 \tilde{T}\sin^2 \frac{\alpha}{2}  R_{S3}\right)  \nonumber \\
   &+& \frac{1}{4} \left(H_0+i A_0\right) y_R \bar{t}_R \left(\sqrt{2} \tilde{X}_{\frac{2}{3}} \sin \alpha 
   R_{A1} -2 \tilde{T}_X \sin^2 \frac{\alpha}{2} R_{S3} \right) \nonumber \\
  &+& \frac{1}{2} \left(H_0-i A_0\right) y_L \bar{b}_L \left(\tilde{B} \sin \alpha 
   Q_{A1}-\tilde{X}_{-\frac{1}{3}} Q_{S2}\right) + h.c.
 \eeq

\section{Mixing with $(3,1)+(1,3)$ top partners}\label{Appendix3}

The calculation of oblique parameters in the bi-doublet mixing scenario is given in~\cite{Cai:2022zqu}.  Here we will show the relevant detail for  the triplet mixing scenario. When the top quark mass is generated by the mixing among  the SM  $(t_L, b_L)$ and $t_R$  and  top partners in $(3,1)+(1,3)$ representations,   the  fermions  can be arranged  into up and down sectors:
\beq
\mathcal{U} \equiv \left( t,  X_{2/3}, Y_{2/3} \right)^T \, \quad  \mathcal{D} \equiv \left( b,  X_{-1/3}, Y_{-1/3} \right)^T
\eeq
\beq
M_{2/3} =  \left( \begin{array}{ccc}
0 & - \frac{y_L Q_{S2}}{\sqrt{2}}  f \sin \alpha  &    \frac{y_L Q_{S2}}{\sqrt{2}}  f \sin \alpha \\
y_{R}  R_{A4} f  \sin^2 \frac{\alpha}{2} &  M & 0 \\
y_{R}  R_{A4} f \cos^2 \frac{\alpha}{2} & 0 & M
\end{array}
\right) 
\eeq

\beq
M_{-1/3} = \left(
 \begin{array}{ccc}
 0 & - y_L  Q_{S2} f \sin \alpha  & y_L   Q_{S2} f \sin \alpha  \\
 0 & M & 0 \\
 0 & 0 & M \\
\end{array}
\right)
\eeq
where we can rescale $ y_L Q_{S2} \to y_L $ and  $ y_R  R_{A4} \to  y_R$ without losing generality.  The up and down masses are diagonalized  by the basis rotation,  i.e. $\Omega^\dag_L M_{2/3} \Omega_R = M_{2/3}^{diag}$ and  $\Omega^{d \dag}_L M_{-1/3} \Omega_R^d = M_{-1/3}^{diag}$. And the rotation matrices  for the triplet scenario  at $\mathcal{O}(\epsilon^2)$ ($\epsilon \equiv \sin \alpha$) are:
\beq
\Omega_L = \left(
\begin{array}{ccc}
 \frac{f^2 \epsilon ^2 \left(-\frac{M^4}{\left(M^2+f^2 y_R^2\right){}^2}-1\right) y_L^2}{4
   M^2}+1 & -\frac{f \epsilon  y_L}{\sqrt{2} M} & \frac{f M \epsilon  y_L}{\sqrt{2}
   \left(M^2+f^2 y_R^2\right)} \\
 \frac{f \epsilon  y_L}{\sqrt{2} M} & 1-\frac{f^2 \epsilon ^2 y_L^2}{4 M^2} & \epsilon ^2
   \left(\frac{1}{4}-\frac{M^2 y_L^2}{2 f^2 y_R^4+2 M^2 y_R^2}\right) \\
 -\frac{f M \epsilon  y_L}{\sqrt{2} \left(M^2+f^2 y_R^2\right)} & \epsilon ^2
   \left(\frac{y_L^2}{2 y_R^2}-\frac{1}{4}\right) & 1-\frac{f^2 M^2 \epsilon ^2 y_L^2}{4
   \left(M^2+f^2 y_R^2\right){}^2} \\
\end{array}
\right)
\eeq
 
\beq
 \Omega_R &=& \left(
\begin{array}{ccc}
 \frac{M}{\sqrt{M^2+f^2 y_R^2}} & 0 & \frac{f y_R}{\sqrt{M^2+f^2 y_R^2}} \\
 0 & 1 & 0 \\
 -\frac{f y_R}{\sqrt{M^2+f^2 y_R^2}} & 0 & \frac{M}{\sqrt{M^2+f^2 y_R^2}} \\
\end{array}
\right) \nonumber \\
&+& \epsilon^2 \left(
\begin{array}{ccc}
 \frac{f^2 M  y_R^2 \left(M^2+ f^2 ( 2 y_L^2+ y_R^2 ) \right)}{4 \left(M^2+f^2
   y_R^2\right)^{5/2}} & \frac{f  y_L^2}{2 M y_R} & -\frac{f M^2  y_R
   \left(M^2+ f^2 (2 y_L^2+ y_R^2 ) \right)}{4 \left(M^2+f^2 y_R^2\right)^{5/2}} \\
 -\frac{\left(M^2+2 f^2 y_L^2\right)}{4 M^2 \sqrt{\frac{M^2}{f^2 y_R^2}+1}} & 0 &
   \frac{M  \left(y_R^2-2 y_L^2\right)}{4 y_R^2 \sqrt{M^2+f^2 y_R^2}} \\
 \frac{M^2  \left(M^2+ f^2 (2 y_L^2+  y_R^2) \right)}{4 \sqrt{\frac{M^2}{f^2 y_R^2}+1}
   \left(M^2+f^2 y_R^2\right)^2} & \frac{y_L^2}{2}  
   \left(\frac{1}{y_R^2}-\frac{ f^2}{M^2}\right)- \frac{1}{4} & \frac{f^2 M  y_R^2
   \left(M^2+  f^2 (2 y_L^2+ y_R^2 ) \right)}{4 \left(M^2+f^2 y_R^2\right)^{5/2}} \\
\end{array}
\right)
\eeq

\beq
\Omega_L^d = \left(
\begin{array}{ccc}
 1-\frac{f^2 \epsilon ^2 y_L^2}{M^2} & 0 & -\frac{\sqrt{2} f \epsilon  y_L}{M} \\
 \frac{f \epsilon  y_L}{M} & \frac{1}{\sqrt{2}} & \frac{M^2-f^2 \epsilon ^2 y_L^2}{\sqrt{2} M^2}
   \\
 -\frac{f \epsilon  y_L}{M} & \frac{1}{\sqrt{2}} & \frac{f^2 \epsilon ^2 y_L^2-M^2}{\sqrt{2} M^2}
   \\
\end{array}
\right) \,, \quad \Omega_R^d = \left(
\begin{array}{ccc}
 1 & 0 & 0 \\
 0 & \frac{1}{\sqrt{2}} & \frac{1}{\sqrt{2}} \\
 0 & \frac{1}{\sqrt{2}} & -\frac{1}{\sqrt{2}} \\
\end{array}
\right)
\eeq
which  observe  $\Omega_{L}^\dag \Omega_L = \Omega_{L}^{d \dag} \Omega_L^d  = \Omega_{R} \Omega_R^\dag = 1 + \mathcal{O}(\epsilon^3)$.  Note that the unitarity  will ensure the contribution to $S,T$ from the rotation effect to be finite. For the triplet mixing,  the fermion masses  are determined to be:
\beq
&& m_t = \frac{f^2 y_L y_R  \sin \alpha }{\sqrt{2} \sqrt{M^2+ f^2 y_R^2}} \,, \quad  m_{X_{2/3}} = M+ \frac{1}{4 M} f^2  y_L^2  \sin^2 \alpha  \nonumber \\
&& m_{Y_{2/3}} = \sqrt{M^2 + f^2 y_R^2} -\frac{f^2 \sin^2 \alpha \left(f^2 y_R^4+M^2 \left(y_R^2 - y_L^2 \right) \right)}{4 \left(f^2
   y_R^2+M^2\right)^{3/2}} \nonumber \\
&& m_{X_{-1/3}} =  M + \frac{1}{M} f^2 y_L^2 \sin^2 \alpha \,,\quad  m_{Y_{-1/3}} = m_{X_{5/3}}= M
\eeq
In the SU(6)/SO(6) model, the basis rotation contribution to oblique parameters is:
\beq
\Delta T_{mix}  &=&   \frac{N_c}{16 \pi s^2_W c^2_W}   \frac{m_t^2}{m^2_{Y_{2/3}}} \Big[  \frac{1}{\sin^2\phi_R \cos^2\phi_R}  \Big( 3 \theta_+ (y_t, y_M)   - 3  \theta_+ (y_b,  y_{M})  -\theta_+ (y_t, y_b) \Big)  \nonumber \\ &+& \frac{\cos^2\phi_R}{\sin^2\phi_R} \Big( \theta_+ (y_{Y_{\frac{2}{3}}} , y_b) -  \theta_+ (y_t, y_{Y_{\frac{2}{3}}})  -\theta_+ (y_t, y_b) \Big) \Big]
\eeq
\beq
\Delta S_{mix} &=& - \frac{N_c}{2 \pi }  \left[   \frac{m_t^2}{m^2_{Y_{2/3}}} \left[ \frac{4}{\sin^2 2 \phi_R }  \Big(  \chi_+ (y_t, y_{X_{\frac{2}{3}}}) +  4 \chi_+ (y_b, y_{Y_{-\frac{1}{3}}})  \Big) +   \frac{\cos^2\phi_R}{\sin^2\phi_R}    \chi_+ (y_t, y_{Y_{\frac{2}{3}}}) \right] \right. \nonumber \\& - &  \left. \Big[\frac{8}{9} \log \frac{y_{X_{-\frac{1}{3}}}}{y_{X_{\frac{5}{3}}}} + \frac{1}{9} \frac{m_t^2}{m^2_{Y_{2/3}}}   \frac{1}{\cos^2 \phi_R \sin^2 \phi_R^2}  \log\frac{y_t}{y_{X_{\frac{2}{3}}}}  + \frac{1}{9} \frac{m_t^2}{m^2_{Y_{2/3}}}  \frac{\cos^2 \phi_R}{\sin^2 \phi_R}  \log\frac{y_t}{y_{Y_{\frac{2}{3}}}}  \right.
\nonumber \\ &+& \left.  \frac{4}{3}  \frac{m_t^2}{m^2_{Y_{2/3}}}   \frac{1}{\cos^2 \phi_R \sin^2 \phi_R^2}  \left(1- \log\frac{y_{Y_{-\frac{1}{3}}}}{y_b}  \right) \Big] \right]
\eeq 
where the first term in the second line is also in $\mathcal{O}( \sin^2 \alpha)$ because  the mass splitting inside the triplet is
\beq
 m_{X_{-1/3}}-m_{X_{5/3}}= \frac{8\cos \phi_R}{ \sin^2 2 \phi_R} \frac{m_t^2}{m^2_{Y_{2/3}}}  \,, \quad  \sin \phi_R = \frac{f y_R}{\sqrt{M^2 + f^2 y_R^2}}\,. 
\eeq
However as the misalignment connects the bi-doublet with the triplets,  for its contribution to $S$ and $T$, we need to include  the  interactions of  all  top partners in $\Psi_A$.  In the triplet mixing scenario,  the custodial symmetry is conserved at $\mathcal{O}(\sin^2 \alpha)$. Hence we derive the corresponding $S$ parameter to be: 
\beq
\Delta S_{mis} &=&   \Delta S_{div} - \frac{N_c }{2 \pi } \frac{\kappa'^2}{8}   \sin^2\alpha  \Big[    \sin^2 \phi_R \chi_+ (y_t, y_M)  + \left( \cos^2 \phi_R + 1 \right)  \chi_+ (y_{Y_{\frac{2}{3}}}, y_M ) \nonumber \\ 
&+& 2 \cos \phi_R \left[  \chi_- (y_{Y_{\frac{2}{3}}}, y_M ) - \psi_- (y_{Y_{\frac{2}{3}}}, y_M ) \right]  \Big]  
\eeq
with the divergent term included in:
\beq
 \Delta S_{div} &=& \frac{N_c }{2 \pi } \Big[ - \frac{\kappa'^2}{16} \sin^2 \alpha  \Big[ \sin^2 \phi_R \left(\frac{1}{3} - \frac{1}{3} \log y_t^2 \right)  +  (\cos^2 \phi_R  + 1 )  \left(\frac{1}{3} - \frac{1}{3} \log y_{Y_{\frac{2}{3}}}^2 \right)  \Big] \nonumber \\
&+&  \left( 24 \sin^2 \frac{\alpha}{2} - \frac{11}{8} \kappa'^2 \sin^2 \alpha \right) \left(\frac{1}{3} - \frac{1}{3} \log y_M^2 \right) + \frac{\kappa'^2}{12} \sin^2 \alpha \nonumber \\ &+& \left( 16 \sin^2 \frac{\alpha}{2} -  \kappa'^2 \sin^2 \alpha \right)  \left( \log\frac{\Lambda^2}{m^2_Z} -\frac{7}{6} \right)   \Big]
\eeq
which  will recover Eq.(\ref{dS1})  for $\sin \phi_R = 0$,  just  like  the bi-doublet mixing scenario.
Note that  the loop functions  we used  in the oblique parameters are defined in~\cite{Lavoura:1992np}:
\beq
\theta_+(y_1, y_2) &=& y_1+y_2 -\frac{2 y_1y_2 \log \left(\frac{y_1}{y_2}\right)}{y_1-y_2}   \\
\theta_-(y_1, y_2) &=& 2 \sqrt{y_1y_2} \left(\frac{(y_1+ y_2) \log
\left(\frac{y_1}{y_2}\right)}{y_1-y_2}-2\right) \\
\chi_+ (y_1, y_2) &=& \frac{\left(3 y_1y_2 (y_1+ y_2) - y_1^3 - y_2^3\right) \log
   \left(\frac{y_1}{y_2}\right)}{3 (y_1- y_2)^3}+\frac{5 \left(y_1^2+ y_2^2\right)-22 y_1 y_2}{9 (y_1-y_2)^2} \\
  \chi_-(y_1, y_2) &=& -\sqrt{y_1y_2} \left(\frac{y_1+ y_2}{6 y_1 y_2}-\frac{y_1+ y_2}{(y_1- y_2)^2}+\frac{2 y_1 y_2 \log \left(\frac{y_1}{y_2}\right)}{(y_1-y_2)^3}\right) \\
  \psi_-(y_1, y_2) &=& - \frac{y_1 + y_2}{6 \sqrt{y_1 y_2}}
\eeq
Another loop function can  appear in the $S$ parameter originating from  a pure rotation effect:
\beq
  \psi_+ (y_a, y_b) &=&  \frac{1}{3} \left( Q_a - Q_b \right) - \frac{1}{3} \left( Q_a + Q_b \right) \log \left(\frac{y_a}{y_b}\right) 
\eeq
that  is  generalized for the VLQ in a non-standard doublet or triplet representation~\cite{Cai:2022zqu}.

\bibliographystyle{JHEP}

\bibliography{CHMtop.bib}

\end{document}